\documentclass[12pt,dvips]{article}
\usepackage{rotating}
\usepackage{axodraw}
\usepackage{epsfig}
\usepackage{color}
\usepackage{cite}

\def\theequation{\arabic{section}.\arabic{equation}}

\newcommand{\imag}{\Im {\rm m}}

\newcommand{\lsim}{\raisebox{-0.13cm}{~\shortstack{$<$ \\[-0.07cm] $\sim$}}~}
\newcommand{\gsim}{\raisebox{-0.13cm}{~\shortstack{$>$ \\[-0.07cm] $\sim$}}~}

\begin{document}

\def\thefootnote{\fnsymbol{footnote}}

\begin{flushright}
{\tt CERN-PH-TH/2010-137}, {\tt MAN/HEP/2010/07 }\\
{\tt arXiv:1006.3087} \\
June 2010
\end{flushright}

\begin{center}
{\bf {\LARGE A Geometric Approach to CP Violation:\\[4mm]
    Applications to the MCPMFV SUSY Model}}
\end{center}

\medskip

\begin{center}{\large
John~Ellis$^a$,
Jae~Sik~Lee$^b$ and
Apostolos~Pilaftsis$^c$}
\end{center}

\begin{center}
{\em $^a$Theory Division, CERN, CH-1211 Geneva 23, Switzerland}\\[0.2cm]
{\em $^b$Physics Division, National Center for Theoretical Sciences, 
Hsinchu, Taiwan 300}\\[0.2cm]
{\em $^c$School of Physics and Astronomy, University of Manchester,}\\
{\em Manchester M13 9PL, United Kingdom}
\end{center}

\bigskip\bigskip

\centerline{\bf ABSTRACT}

\noindent  
We  analyze the constraints  imposed by  experimental upper  limits on
electric dipole moments (EDMs)  within the Maximally CP- and Minimally
Flavour-Violating  (MCPMFV) version  of  the MSSM.   Since the  MCPMFV
scenario has  6 non-standard CP-violating  phases, in addition  to the
CP-odd QCD  vacuum phase  $\theta_{\rm QCD}$, cancellations  may occur
among the CP-violating contributions to the three measured EDMs, those
of  $^{205}$Tl,   the  neutron   and  $^{199}$Hg,  leaving   open  the
possibility  of  relatively large  values  of  the other  CP-violating
observables.   We  develop a  novel  geometric  method  that uses  the
small-phase approximation as a  starting point, takes the existing EDM
constraints into  account, and  enables us to  find maximal  values of
other CP-violating observables,  such as the EDMs of  the Deuteron and
muon, the  CP-violating asymmetry in $b  \to s \gamma$  decay, and the
$B_s$ mixing  phase. We apply  this geometric method to  provide upper
limits on  these observables within  specific benchmark supersymmetric
scenarios, including extensions that allow for a non-zero $\theta_{\rm
QCD}$.

\newpage

\section{Introduction}

The  non-observation of  electric dipole  moments (EDMs)  for Thallium
($^{205}{\rm               Tl}$)~\cite{Regan:2002ta},              the
neutron~($n$)~\cite{Baker:2006ts},     and     Mercury    ($^{199}{\rm
Hg}$)~\cite{Romalis:2000mg,Griffith:2009zz} provide  very tight bounds
on    possible    new    sources    of   CP violation    beyond    the
Cabibbo--Kobayashi--Maskawa  (CKM) phase  of the  Standard Model~(SM).
{\it Prima  facie}, these bounds suggest that  any CP-violating phases
associated with new physics at the  TeV scale are very small, posing a
challenge to scenarios of new TeV-scale physics, such as supersymmetry
(SUSY)~\cite{SUSY},  that contain  many  potential new  sources of  CP
violation.  The EDM challenge is compounded by the excellent agreement
of  present  experiments  with  the  CKM  model~\cite{CKM},  providing
important complementary  constraints on  the flavour structure  of any
new TeV-scale physics, as well as  on its role in CP violation. On the
other  hand, the  baryon  asymmetry  in the  Universe  (BAU) could  be
explained by TeV-scale physics, if it has substantial CP violation and
realizes   a  sufficiently   strong   first-order  electroweak   phase
transition in the early Universe~\cite{BAU}.

This tension  between a TeV-scale origin  of the BAU and  EDMs will be
explored both directly  and indirectly in the LHC  era. Experiments at
the  LHC, notably  LHCb~\cite{LHCb}, will  soon be  giving  direct new
information  about flavour  and  CP  violation at  the  TeV scale.  In
parallel to these direct explorations  at the LHC, a new generation of
precision low-energy experiments will play an important indirect role.
These  new precision  experiments  will place  much stronger  indirect
constraints  on the  possible CP  and flavour  structure of  models of
TeV-scale physics.   New experiments on  the neutron EDM  are underway
and,  if the  proposed experiment  searching for  a  Deuteron ($^2{\rm
H}^+$)           EDM           achieves           its           design
sensitivity~\cite{Semertzidis:2003iq,OMS},   it   will   improve   the
existing bounds  on possible CP-violating  chromoelectric operators by
orders of magnitude~\cite{Lebedev:2004va}.

In  this  paper   we  introduce  a  new  geometric   approach  to  the
incorporation of  EDM constraints on CP-violating  models, showing how
the  maximal  values of  unmeasured  CP-violating  observables may  be
estimated  in  a systematic  and  reliable  way.   We illustrate  this
approach in the  context of SUSY, regarding it as  an archetype of the
TeV-scale  models that are  (potentially) embarrassed  by experimental
constraints  on  flavour and  CP  violation.   However, our  geometric
approach could easily be applied  to other models, and indeed to other
classes of observables besides those that violate CP.

For illustration, we work  within the minimal supersymmetric extension
of  the Standard  Model (MSSM),  with SUSY  broken softly  at  the TeV
scale. We assume a generic  version of this model with minimal flavour
violation (MFV), driven by  the fermion Yukawa couplings. As discussed
in~\cite{Ellis:2007kb}          and         elaborated         further
in~\cite{Ellis:2009di,RefMFV},   this  model   has  a   total   of  19
parameters, of which  6 violate CP.  In the  convention where both the
superpotential  Higgs-mixing parameter $\mu$  and the  respective soft
SUSY-breaking  Higgs-mass term $B\mu$  are real,  these are  the three
phases of  the soft SUSY-breaking  gaugino masses $M_{1,2,3}$  and the
three  phases of the  trilinear SUSY-breaking  parameters $A_{d,u,e}$.
This  model was  called in~\cite{Ellis:2007kb}  the Maximally  CP- and
Minimally  Flavour-Violating (MCPMFV)  scenario  of the  MSSM, or  the
MCPMFV SUSY model  in short.  In addition to the  6 MCPMFV phases, one
may  consider the  CP-violating QCD  vacuum phase  $\theta_{\rm QCD}$.
The  specific question  we study  in  this paper  is how  to find  the
maximum  value of  some  other CP-violating  observable,  such as  the
CP-violating asymmetry  in $b \to  s \gamma$ decay, $A_{\rm  CP}$, the
phase in  $B_s$ mixing, $\phi_{B_s}$, or  some other EDM  that has not
yet been  measured (accurately),  such as those  for the  Deuteron and
muon, while implementing the available EDM constraints.

In order  to demonstrate the  principle of our geometric  approach, we
first consider a toy warm-up  problem involving a single constraint in
a  theory with  three  parameters.  We  characterize  the subspace  of
parameters satisfying this constraint in the linear approximation, and
then  show  how  to  identify  the direction  in  this  subspace  that
maximizes any  given observable  $O$.  The generalization  to multiple
constraints  in higher-dimensional  spaces  follows similar  geometric
principles, that may be described in the language of exterior products
of differential  forms~\cite{SMC}. This geometric  optimization method
of  the  so-called   ``cancellation  mechanism''~\cite{IN}  may  sound
complicated, but its  numerical implementation is straightforward. 
We emphasize that  our geometric  approach differs
in principle from  the naive scan  method that is usually  adopted in
the  literature~\cite{RefSCAN}.
The geometric  method proposed  here provides an accurate parametric
determination of the optimal cancellation regions where any
given  physical observable is maximized in  the linear  approximation. Hence,  our
geometric approach is exact,  efficient and less computationally-intensive
than a naive scan of a multi-dimensional space.

In the application of this approach to the MCPMFV SUSY model, we first
select some  benchmark points in the CP-conserving  restriction of the
model. We  then evaluate the  dependences of the  relevant constraints
(the EDMs  of $^{205}{\rm Tl}$,  the neutron and $^{199}{\rm  Hg}$) on
the six  CP-violating phases  of the full  MCPMFV model in  the linear
approximation, as  well as the linear dependences  of the CP-violating
observables of interest (the Deuteron  and muon EDMs, $A_{\rm CP}$ and
$\phi_{B_s}$).   We then  demonstrate  numerically how  each of  these
observables may be maximized  in the linear approximation, taking into
account the existing  EDM constraints.  We note that,  at any specific
benchmark  point,  the  values  of all  CP-violating  observables  are
bounded  in magnitude,  since the  ranges of  the  CP-violating phases
$\phi_i$ are all compact: $\phi_i \in  [0, 2 \pi)$.  We also note that
our   approach   is    only   approximate   for   large   CP-violating
phases. Nevertheless, the linear expansions give good estimates of the
true maximal values of the CP-violating observables.

We  find  that  the  linear  approximations  to  the  EDMs  and  other
CP-violating observables in the neighbourhoods of the MCPMFV benchmark
points  we  study are  quite  accurate  for  CP-violating angles  with
magnitudes up to several tens of degrees. We confirm that the EDM-free
directions  in   parameter  space  constructed   using  our  geometric
construction yield  values of the other  CP-violating observables that
are  larger than those  possible (in  the linear  approximation) along
other  directions  in the  space  of  CP-violating  phases. Along  the
optimal EDM-free directions,  we find that values of  the Deuteron EDM
an  order  of  magnitude  larger  than  the  prospective  experimental
sensitivity  may  be attained  for  acceptable  values  of the  MCPMFV
phases, and almost  an order of magnitude larger  still if the optimal
geometric  construction is  extended to  include the  CP-violating QCD
vacuum phase $\theta_{\rm  QCD}$.  On the other hand,  we find maximal
values  of  the  muon  EDM  that are  below  the  likely  experimental
sensitivity in both the scenarios  with and without the QCD phase.  In
the case of $A_{\rm CP}$, we find  values as large as 2\% in the large
$\tan\beta$  scenario. Given  the constraint  from  $B(b\to s\gamma)$,
however, $A_{\rm CP}$ cannot exceed  the 0.1\% level and so remains too
small  to   be  observed.   Finally,  the   $B_s$-meson  mixing  phase
$\phi_{B_s}$ turns out  to be close to the small SM  value in both the
scenarios studied.

The layout  of the paper is  as follows. Section~2  presents our novel
geometric approach,  starting with the  toy three-dimensional exercise
and continuing to  the six-dimensional case of the  MCPMFV SUSY model.
The  implementation of  this  approach for  a  selection of  benchmark
points  is  described in  Section~3,  and  our  numerical results  for
$A_{\rm  CP}$,  $\phi_{B_s}$  and  the  Deuteron  and  muon  EDMs  are
presented in  Section~4.  Section~5 extends the previous discussion to
the seven-dimensional case including $\theta_{\rm QCD}$.
Finally, Section~6  presents our conclusions
and some suggestions for future work.


\setcounter{equation}{0}
\section{Optimal EDM-Free Directions}\label{sec:geometry}

The  current experimental upper  bounds on  the Thallium,  neutron and
Mercury  EDMs   put  very  strict  constraints   on  the  CP-violating
parameters of the theory, such as CP-odd phases.  However, they do not
constrain  all possible  combinations  of the  CP-violating phases  in
models  with many  such phases,  such as  the MSSM.   It  is therefore
important  to develop  a  powerful approach  for  finding the  optimal
choice  of  CP-odd   phases  which  maximize  the  size   of  a  given
CP-violating  observable  $O$,  while  remaining compatible  with  the
present  EDM constraints.  Examples  of such  CP-violating observables
for which the present experimental sensitivities are likely soon to be
improved significantly include the Deuteron  and muon EDMs, the CP
asymmetry in~$b \to s \gamma$, and the phase in $B_s$ mixing.

In this paper, we propose  a geometric approach to the maximization of
such  observables,  which  may  also  be applied  to  other  analogous
problems.    For   illustration,    we   first   consider   a   simple
three-dimensional  (3D) example  for  a theory  which  has only  three
physical CP-odd  phases.  Working  in the 3D  vector space  defined by
these  three CP-odd  phases, we  show how  to  construct geometrically
optimal  `EDM-free' directions  in the  small-phase  approximation. We
then generalize this approach to  theories with more than three CP-odd
phases, such as the MCPMFV SUSY  model with its six new CP-odd phases,
optionally including  the QCD vacuum phase  $\theta_{\rm QCD}$.  Using
simple ideas  from the  calculus of differential  forms~\cite{SMC} for
such a  higher-dimensional setup,  we are able  to derive  the optimal
EDM-free directions for any  given CP-violating observable $O$ that we
wish to study.

\subsection{A Simple 3D Example}

For the sake of geometric familiarity, we first consider the simple 3D
example  of   a  theory  with  just  three   physical  CP-odd  phases,
represented  by the 3D  phase vector  {\boldmath $\Phi$}~$  = (\Phi_1,
\Phi_2, \Phi_3)$.  For  illustration, we assume that we  have a single
very stringent EDM constraint, which  we denote by $E$.  In the region
of     small      phases,     e.g.,     for      {\boldmath     $|\Phi
|$}~$\stackrel{<}{{}_\sim} \pi/6$,  we seek  to maximize the  value of
some  specific  CP-violating   observable  $O$,  under  the  condition
$E=0$. Both the CP-violating observable $O$ and the EDM constraint $E$
are  functions of  {\boldmath  $\Phi$}, i.e.,  $O =  O(\mbox{\boldmath
$\Phi$})$  and $E  = E(\mbox{\boldmath  $\Phi$})$, and  vanish  in the
limit  of vanishing  CP-odd  phases:  $O ({\bf  0})  =0$, $E({\bf  0})
=0$~\footnote{Our approach can easily  be extended to cases where some
of  the  phases $\Phi_{1,2,3}$  approach  another CP-conserving  point
different from zero,  e.g., $\Phi_{1,2,3} \to \pm \pi$,  in which case
the CP-odd  phase vector {\boldmath $\Phi$}  represents the difference
from this CP-conserving point.}.

\begin{figure}[t]
\begin{center}
\begin{picture}(200,180)(0,50)
\SetWidth{0.9}

\LongArrow(100,100)(200,80)\Text(205,80)[l]{\boldmath$\widehat{\Phi}_2$}
\LongArrow(100,100)(100,185)\Text(100,190)[b]{\boldmath $\widehat{\Phi}_3$}
\LongArrow(100,100)(30,60)\Text(25,60)[r]{\boldmath $\widehat{\Phi}_1$}
\Line(100,100)(170,140)

\LongArrow(100,100)(130,150)\Text(125,155)[lb]{${\bf E} \equiv \nabla E$}
\LongArrow(100,100)(70,150)\Text(80,155)[rb]{${\bf O} \equiv \nabla O$}
\LongArrow(100,100)(70,130)\Text(70,125)[rt]{\boldmath $\Phi^*$}

\Line(30,110)(100,150)
\DashLine(85,50)(150,87){3}
\Line(30,110)(57,80)\DashLine(57,80)(85,50){3}
\Line(100,150)(127,120)\DashLine(127,120)(155,90){3}

\end{picture} 
\end{center}
\caption{\it Geometric construction  of the optimal EDM-free direction
  in the small-phase approximation  for a CP-violating observable $O$,
  subject to  the EDM  constraint $E=0$, in  a theory with  three physical
  CP-odd  phases  $\Phi_{1,2,3}$. The  optimal  EDM-free direction  is
  given by  the CP-odd phase  vector {\boldmath $\Phi^*$}, which  is the
  intersection  of the  indicated  plane perpendicular  to the  vector
  ${\bf E}$ with the plane defined  by the vectors ${\bf O}$ and ${\bf
  E}$.}
\label{fig:geometry}
\end{figure}
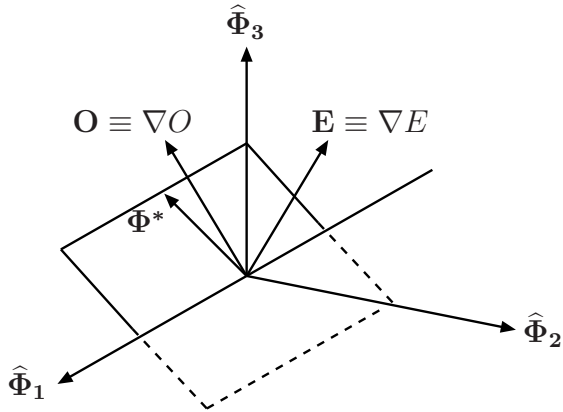

We make Taylor expansions of $O$ and $E$ in terms of the small phases, and
keep only the linear terms in these expansions:
\begin{equation}
  \label{defOE}
O\ =\ \mbox{\boldmath $\Phi$} \cdot {\bf O}\;, \qquad
E\ =\ \mbox{\boldmath $\Phi$} \cdot {\bf E}\;,
\end{equation}
where  we have  defined ${\bf  O} \equiv  \nabla O$,  ${\bf  E} \equiv
\nabla    E$, and    $\nabla    \equiv    (\partial/\partial\phi_1,
\partial/\partial\phi_2,   \partial/\partial\phi_3)$.   The  condition
$E=0$ requires that the phase  vector {\boldmath $\Phi$} should lie in
the plane  orthogonal to  ${\bf E}$. In order to  maximize the
value of $O$, the phase  vector {\boldmath $\Phi$} should lie along the
intersection of the  plane spanned by the vectors ${\bf
  O}$  and ${\bf  E}$ with  this plane  perpendicular to  ${\bf E}$,
 as  represented  schematically    in
Fig.~\ref{fig:geometry}. Up to an overall sign, the solution is
unique and is given by the double cross-product:
\begin{equation}
  \label{EDMfree}
\mbox{\boldmath $\Phi^*$}\ =\ {\bf E}\times 
                                  \Big( {\bf O} \times {\bf E} \Big)\; .
\end{equation}
Evidently, the  condition $E=0$ is satisfied  by construction
in the small-phase approximation, i.e., $E
\equiv \mbox{\boldmath  $\Phi^*$} \cdot {\bf  E} = 0$. 
In this way, we can construct  unambiguously the optimal EDM-free 
direction {\boldmath   $\Phi^*$}.

It is straightforward to obtain  the maximum value of the CP-violating
observable~$O$. In the small-phase approximation, this is simply given by
\begin{equation}
 \label{Ooptimal}
O\ =\ \phi^*\; \mbox{\boldmath $\widehat{\Phi}^*$} \cdot {\bf O}\ =\
\pm\, \phi^*\; \sqrt{|{\bf O}|^2\: -\: ({\bf O}\cdot {\bf \widehat{E}})^2}\ ,
\end{equation}
where $\phi^*  \equiv$~{\boldmath $|\Phi^*|$} and
the carets on the  vectors indicate unit-norm vectors. 
In the small-phase approximation,
the  largest possible  value  for the  CP-violating observable~$O$  is
obtained when ${\bf  E}$ is perpendicular to ${\bf O}$,
in which case we have
\begin{equation}
  \label{Omax}
O_{\rm max}\ =\ \pm\; \phi^*\, |{\bf O}|\ .
\end{equation}
On the  other extreme, if  ${\bf E}$ happens  to be parallel  to ${\bf
O}$, then  the CP-violating observable~$O$ vanishes, i.e., $O  = 0$,
in the limit that $E = 0$. Note
that there is a twofold degeneracy in the optimal value of
the  CP-violating  observable  $O$,  i.e., our  geometric  construction
leaves $O$ undetermined  up to an overall sign.  However, this twofold
degeneracy is  a consequence of the  linear small-phase approximation,
where     quadratic     and     higher-order     derivative     terms
in a Taylor series expansion, e.g., $\nabla_i\nabla_j E$, 
were  neglected. These terms break this twofold degeneracy
in general, and we  include them all in  our  numerical analysis.

\subsection{A 6D Example: the MCPMFV SUSY Model}

The above geometric construction for the 3D example can be generalized
to theories  with more  than three CP-odd  phases, such as  the MCPMFV
SUSY model,  which has six phases~\footnote{We  extend this discussion
later  to include the  QCD phase  $\theta_{\rm QCD}$.},  and including
more than  one EDM constraint. In the  case of such a  theory with six
CP-odd phases, {\boldmath $\Phi$} is now a 6D phase vector, subject to
three EDM constraints denoted  by $E^{a,b,c}= 0$, corresponding to the
non-observation of the Thallium, neutron and Mercury EDMs.  As before,
the  task  is  to   maximize  a  given  CP-violating  observable  $O$,
now satisfying simultaneously the three conditions: $E^a = E^b = E^c = 0$.

The  generalization  of  the   differential  operator  $\nabla$  to  6
dimensions is obvious, and it may be used to obtain in the small-phase
approximation  the   four  6D  vectors:  ${\bf   E}^{a,b,c}  =  \nabla
E^{a,b,c}$ and  ${\bf O} = \nabla  O$. For simplicity,  we assume that
the  four  vectors  ${\bf  E}^{a,b,c}$  and  ${\bf  O}$  are  linearly
independent,  i.e.,   there  are  no  degeneracies   between  the  EDM
constraints and  the observable $O$:  we return to this  issue towards
the end of this Section.

Given the above assumptions, the analogue of the single vector ${\bf E}$ 
in the 3D example is the triple exterior product
\begin{equation}
  \label{A3form}
A_{\alpha\beta\gamma}\ =\ E^a_{[\alpha}\,E^b_{\beta}\,E^c_{\gamma ]}\; ,
\end{equation}
where the Greek indices label the  components of the vectors in the 6D
space, i.e., $\alpha ,\ \beta ,\  \gamma\ =\ 1,2,\dots, 6$.  The square
brackets  on  the  RHS  of~(\ref{A3form})  indicate  that  the  tensor
$A_{\alpha\beta\gamma}$  is  obtained  by fully  antisymmetrizing  the
vectors  $E^a_\alpha$,  $E^b_\beta$ and  $E^c_\gamma$  in the  indices
$\alpha   ,    \beta   ,   \gamma$,    i.e., $A_{\alpha\beta\gamma}   =
-A_{\beta\alpha\gamma}  = -  A_{\alpha\gamma\beta}$  etc. Borrowing  a
term from the  calculus of differential forms, $A_{\alpha\beta\gamma}$
is a 3-form.

Correspondingly,  the analogue  of the  direction ${\bf  O}\times {\bf
  E}$, which determines  the normal to the plane  defined by ${\bf O}$
and ${\bf E}$ in the 3D example described above, is the 2-form
\begin{equation}
  \label{B2form}
B_{\mu\nu}\ =\ \varepsilon_{\mu\nu\lambda\rho\sigma\tau}\, O_\lambda\,
E^a_\rho\, E^b_\sigma\, E^c_\tau\; ,
\end{equation}
where    summation   over    repeated   indices    is    implied   and
$\varepsilon_{\mu\nu\lambda\rho\sigma\tau}$ is  the usual Levi--Civita
tensor generalized  to 6D.  In the language of differential forms, 
$B_{\mu\nu}$ is, up to an  irrelevant overall
factor,  the  Hodge-dual product  between the  1-form
$O_\lambda$, representing the  CP-violating observable, and the 3-form
$A_{\alpha\beta\gamma}$.

The  components  $\Phi^*_\alpha$  of  the optimal  EDM-free  direction
maximizing $O$ can  now be obtained from the  Hodge-dual product of
the 3-form  $A_{\beta\gamma\delta}$ and the  2-form $B_{\mu\nu}$.
Explicitly,
\begin{equation}
  \label{Phi6D}
\Phi^*_\alpha\ =\ {\cal N}\,
\varepsilon_{\alpha\beta\gamma\delta\mu\nu}\,
A_{\beta\gamma\delta}\; B_{\mu\nu}\ =\ 
{\cal N}\; \varepsilon_{\alpha\beta\gamma\delta\mu\nu}\,
\varepsilon_{\mu\nu\lambda\rho\sigma\tau}\,
 E^a_\beta\, E^b_\gamma\, E^c_\delta\; O_\lambda\, E^a_\rho\, E^b_\sigma\, E^c_\tau\ ,
\end{equation}
where we have included  an unknown overall normalization factor ${\cal
  N}$.  By  construction, the 6D phase vector  {\boldmath $\Phi^*$} is
orthogonal to the three vectors ${\bf E}^{a,b,c}$, thus satisfying the
desired EDM  constraints, $E^a =  E^b = E^c  = 0$, in  the small-phase
approximation.    We observe  that   the   magnitude  $\phi^*   \equiv
$~{\boldmath  $|\Phi^*|$},  and hence  the  overall normalization  factor
${\cal  N}$, can only  be determined  by a  numerical analysis  of the
actual experimental limits on the three EDMs.  As in the 3D example,
the maximum allowed value of the CP-violating observable $O$ is given
in the small-phase approximation by
\begin{equation}
  \label{Ooptimal6}
O \ =\ \phi^*\; \widehat{\Phi}^*_\kappa\,
O_\kappa\,\ =\ \pm\; {\cal N}\ \Big|\varepsilon_{\mu\nu\alpha\beta\gamma\delta}\,
  \varepsilon_{\mu\nu\lambda\rho\sigma\tau}\, O_\alpha\, O_\lambda\,
  E^a_\beta\, E^b_\gamma\, E^c_\delta\, E^a_\rho\, E^b_\sigma\,
  E^c_\tau\,\Big|\;,
\end{equation}
where the caret denotes the components  of a unit-norm vector. As in the
3D case,  quadratic and higher-order derivative terms  with respect to
the CP-odd phases  will generically prefer a  particular sign for the
optimal value of $O$~\footnote{In 
some circumstances, the small-phase approximation may even break down
in such a way that the maximal value of $O$ is obtained for a value of
$\phi^*$ less than the maximal value allowed by $E$.}.

As a consistency  check of our geometric construction,  one may verify
that  in the  small-angle  approximation the  largest possible  value,
$O_{\rm max} = \pm\, \phi^* |{\bf  O}|$, is obtained when all the four
6D vectors, ${\bf O}$  and ${\bf E}^{a,b,c}$, are mutually orthogonal,
exactly  as  in  the 3D  case.   For  example,  if the  vectors  ${\bf
E}^{a,b,c}$  have  non-zero  components  only   in  the  4,  5  and  6
directions,  respectively, and  the CP-violating  observable $O$  is a
reduced  3-vector living  in the  1, 2  and 3  coordinates, it  is not
difficult  to check  that the  expression~(\ref{Ooptimal6})  gives the
largest possible value $O_{\rm max}$, as expected.

The above  geometric construction can  be extended to  include further
EDM  or   other  strict   CP-violating  constraints  on   the  theory.
Specifically, in the MCPMFV SUSY model, we can afford to have at least
two  more  linearly-independent   constraints  coming  from  different
experiments and still be able to potentially get one large combination
of CP-odd phases.   If $E^{d,e} = 0$ are  these two extra constraints,
the  maximum allowed value  for the  CP-violating observable  $O$ will
then be
\begin{equation}
  \label{Otriple}
O\ =\ \pm\ {\cal N}\, \Big|\varepsilon_{\mu\nu\alpha\beta\gamma\delta}\
 O_\mu\, E^a_\nu\, E^b_\alpha\, E^c_\beta\, 
                          E^d_\gamma\, E^e_\delta \Big|\; ,
\end{equation}
where  the normalization  ${\cal N}$  depends  on the  actual size  of
$\phi^*$.  Equation~(\ref{Otriple}) is nothing else than the simple
generalization of the well-known triple-product  to  6D.  

We can  also allow for the  possible presence of a  non-zero strong CP
phase   $\theta_{\rm  QCD}$  in   the  theory,   in  which   case  the
corresponding   CP-odd  phase   vector   {\boldmath  $\Phi$}   becomes
seven-dimensional (7D) in  the MCPMFV SUSY model.  In  this case, four
additional  linearly-independent  EDMs  or other  strict  CP-violating
constraints would be needed, in  addition to the three limits from the
present EDM  experiments, in order to  span fully the  CP-odd space of
the MCPMFV  model, and so  constrain the norm  of the 7D  CP-odd phase
vector~{\boldmath $\Phi$} as discussed in Section~5.

In our geometric  construction, an important role is played by the degree of
degeneracy, or alignment,  between pairs of observables, e.g., between $O$
and the $E^a$. For this  purpose, it  is interesting  to know  the cosine
$C_{{\bf  O,E}^a}$  of the  relative  angle  between their  corresponding
vectors ${\bf O}$ and ${\bf E}^a$, i.e.,
\begin{equation}
  \label{Cosine}
C_{{\bf O,E}^a}\ =\ \frac{ {\bf O}\cdot {\bf E}^a}{|{\bf O}|\, |{\bf E}^a|}\ .
\end{equation}
If $C_{{\bf  O,E}^a} =  \pm 1$, we have perfect
alignment of the observables $O$  and~$E^a$. In such a case, if
$E^a$ vanishes, then so does $O$. On the other hand, if $C_{\bf O,E^a}
= 0$, the two observables $O$  and $E^a$ are orthogonal, and hence
can  vary independently of  each other.  Using~(\ref{Cosine}), one
can obtain  an upper bound for  the optimal value  of the CP-violating
observable $O$ in the small-phase approximation:
\begin{equation}
  \label{Oapprox}
|O|\ \le\ 
\phi^*\; \Big[\, 1\: -\: {\rm Max}\,\Big(C^2_{{\bf
  O,E}^x}\Big)\Big]^{1/2}\; |{\bf O}|\; ,
\end{equation}
where the index $x$ labels the vectors ${\bf E}^{a,b,c,\dots}$ related
to    the    different    EDM    constraints.    Notice    that    the
inequality~(\ref{Oapprox}) becomes an exact equality in the 3D example
which            we           discussed            before           in
Section~\ref{sec:geometry}.1~[cf.~(\ref{Ooptimal})].

Finally, after completing the description of this geometric approach,
we note that it could be applied to many
other phenomenological problems where there are $M$ constraints
on a theory with $N > M$ parameters, and one wishes to determine
the maximum value of some other observable quantity.

\setcounter{equation}{0}
\section{Numerical Illustrations}\label{sec:numerix}

In  this Section we  illustrate the  geometric approach  introduced in
Section~\ref{sec:geometry} by constructing optimal EDM-free directions
for some specific benchmark  scenarios.  For this purpose, we consider
CP-violating variants of a typical CMSSM scenario with
\begin{eqnarray}
&&\left|M_{1,2,3}\right|=250~~{\rm GeV}\,, \nonumber \\
&&M^2_{H_u}=M^2_{H_d}=\widetilde{M}^2_Q=\widetilde{M}^2_U=\widetilde{M}^2_D
=\widetilde{M}^2_L=\widetilde{M}^2_E=(100~~{\rm GeV})^2\,, \nonumber \\
&&\left|A_u\right|=\left|A_d\right|=\left|A_e\right|=100~~{\rm GeV}\,,
\label{eq:cpsps1a}
\end{eqnarray}
at the GUT scale, introducing non-zero CP-violating
phases and varying  $\tan\beta\,(M_{\rm SUSY})$.
We adopt the convention that $\Phi_\mu=0^\circ$, and
we vary independently the following six MCPMFV phases at the GUT scale: 
$\Phi_1$, $\Phi_2$, $\Phi_{3}$,
$\Phi_{A_u}$, $\Phi_{A_d}$, and $\Phi_{A_e}$.
We note that the $\Phi_{1,2,3}$, and the $\mu$  parameter, $\Phi_\mu$, are unchanged by
the  RG evolution at  the one-loop  level, whereas  the phases  of the
trilinear  couplings   ${\bf  A}_{u,d,e}$  at  low   scales  could  be
significantly different  from the values  specified at the  GUT scale.
This scenario becomes  the SPS1a point~\cite{SPS} when $\tan\beta=10$,
$\Phi_{1,2,3}=0^\circ$    and   $\Phi_{A_u,A_d,A_e}=180^\circ$.    Our
calculations of the EDMs include the two-loop diagrams mediated by the
$\gamma$-$H^\pm$-$W^\mp$ and $\gamma$-$W^\pm$-$W^\mp$ couplings, which
are   summarized  in   the   Appendix.   Unlike~\cite{Li:2008kz},   we
incorporate the $\Phi_3$ dependence  induced by gluino exchange in the
one-loop $H^\pm$-$u$-$d$ coupling, which becomes relevant
in the region of large $\tan\beta$.

In order to analyze this  scenario, we first make Taylor expansions of
the following five EDMs and two CP-violating observables:
\begin{eqnarray}
&&
d_{\rm Tl}/d_{\rm Tl}^{\rm EXP}\,, \ \
d_{\rm n}/d_{\rm n}^{\rm EXP}\,, \ \
d_{\rm Hg}/d_{\rm Hg}^{\rm EXP}\,,  \ \
d_{\rm D}/d_{\rm D}^{\rm EXP}\,, \ \
d_{\mu}/d_{\mu}^{\rm EXP}\,, \\
&&
A_{\rm CP}(b\to s\gamma)[\%]\,, \ \
\phi_{B_s}\equiv {\rm Arg}\left(
\langle\bar{B}^0_s|{\cal H}_{\rm eff}^{\Delta B=2}|B^0_s\rangle_{\rm SUSY}\right)[^\circ]\,,
\nonumber
\end{eqnarray}
where we choose the following normalization factors
\begin{eqnarray}
&&
d_{\rm Tl}^{\rm EXP}=9\times 10^{-25}\, e\,{\rm cm}\,, \ \
d_{\rm n}^{\rm EXP}=3\times 10^{-26}\, e\,{\rm cm}\,, \ \
d_{\rm Hg}^{\rm EXP}=3.1\times 10^{-29}\, e\,{\rm cm}\,, \nonumber \\
&&
d_{\rm D}^{\rm EXP}=3\times 10^{-27}\, e\,{\rm cm}\,, \ \
d_{\mu}^{\rm EXP}=1\times 10^{-24}\, e\,{\rm cm}.
\end{eqnarray}
In the cases of the EDMs of Thallium~\cite{Regan:2002ta},
the neutron~\cite{Baker:2006ts}, and Mercury~\cite{Romalis:2000mg,Griffith:2009zz},
we use the current experimental limits for normalization, and
for the EDMs of Deuteron and muon we use the sensitivities 
projected in~\cite{Semertzidis:2003iq,Semertzidis:1999kv}.
The observables $A_{\rm CP}$ and $\phi_{B_s}$ are
measured in percent and degrees, respectively.

In order to obtain the vectors representing the EDM constraints and the observables
in the six-dimensional CP-phase space, namely
${\bf E}=\nabla E$ and ${\bf O}=\nabla O$ where
\begin{equation}
\nabla_\alpha    \equiv    (\partial/\partial\Phi_1,
\partial/\partial\Phi_2,   \partial/\partial\Phi_3,
\partial/\partial\Phi_{A_u}, \partial/\partial\Phi_{A_d},\partial/\partial\Phi_{A_e})\,,
\end{equation}
we   calculate    the   EDMs    and   the   observables    in   ranges
$\Delta\Phi=\pm10^\circ$   around  each  chosen   CP-conserving  point
${\widetilde\varphi}_\alpha$, varying independently each of the six CP
phases, and performing a quadratic fit in each case.
\begin{figure}[!t]
\begin{center}
{\epsfig{figure=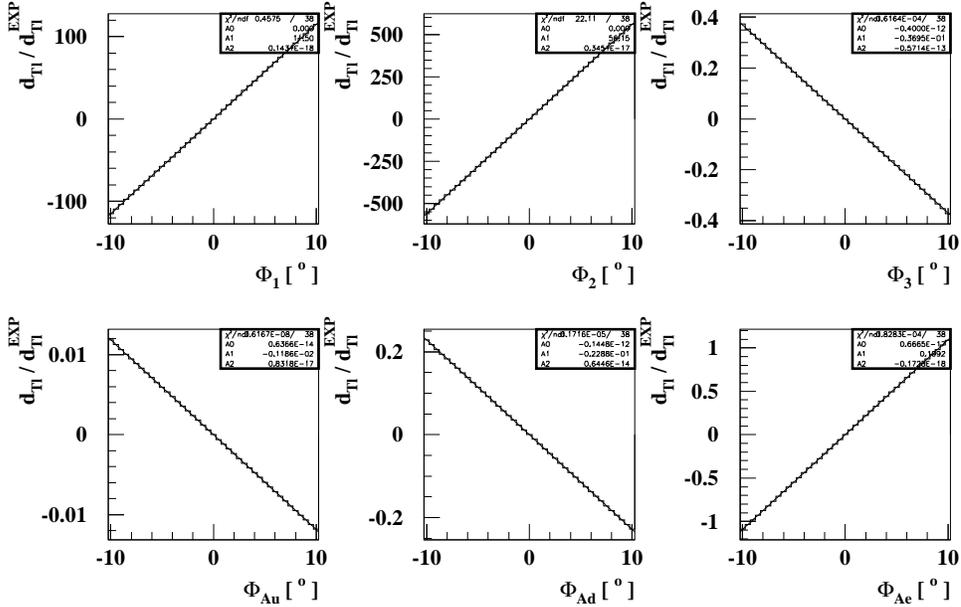,height=14.0cm,width=14.0cm}}
\end{center}
\vspace{-5.0cm}
\caption{\it The quadratic fit to the
Thallium EDM that is used to obtain the 6D vector 
${\bf E}^{\,d_{\rm Tl}}\equiv\nabla (d_{\rm Tl}/d_{\rm Tl}^{\rm EXP})$
in an expansion around $\widetilde\varphi_\alpha=0^\circ$ for 
the scenario~(\ref{eq:cpsps1a}) with $\tan\beta=40$.
} 
\label{fig:dtlfit.0}
\end{figure}
As an example, Fig.~\ref{fig:dtlfit.0} shows 
a quadratic fit to the Thallium EDM for the scenario under consideration
taking $\tan\beta=40$. We observe that
the CP-violating phase dependence is in fact linear in the region considered,
to a very good approximation.
\begin{figure}[!t]
\begin{center}
{\epsfig{figure=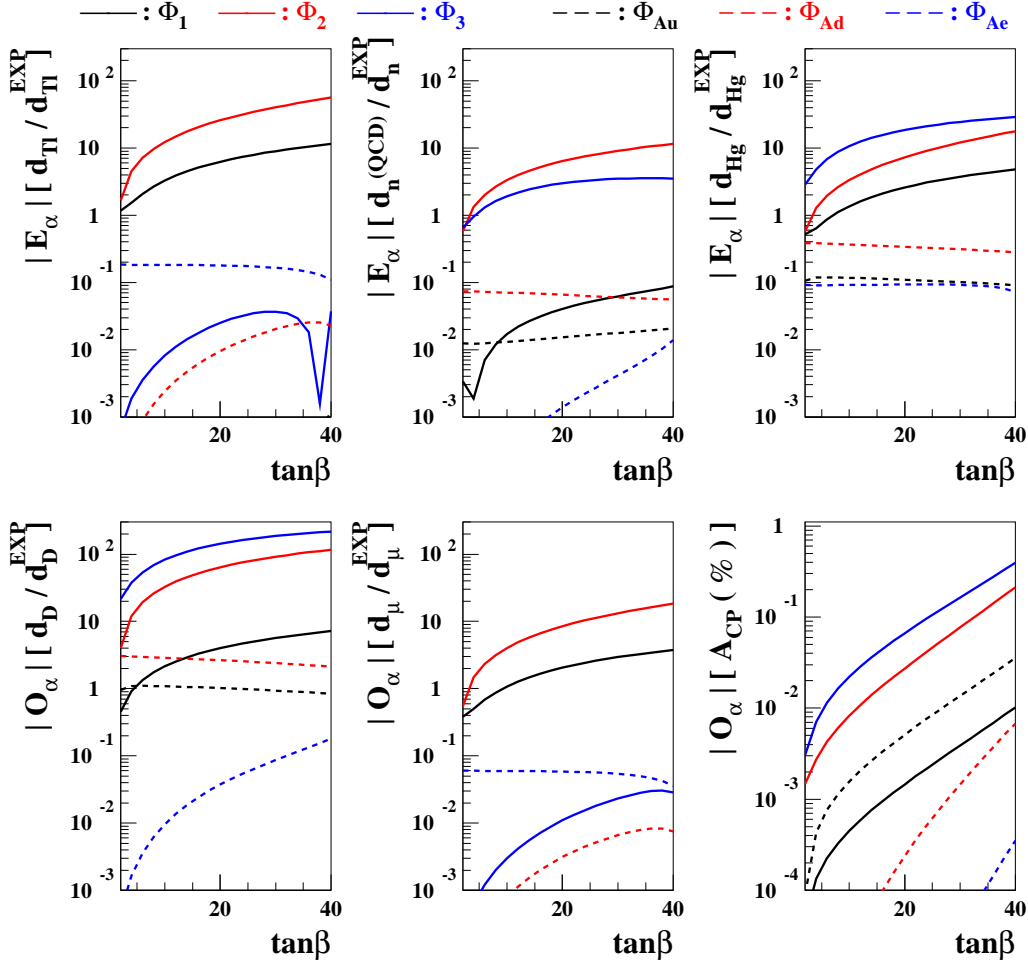,height=14.0cm,width=14.0cm}}
\end{center}
\vspace{-1.0cm}
\caption{\it The absolute values of the components of the three 6D vectors 
representing the present EDM constraints
(upper), and those of the three 6D vectors representing other CP-violating observables (lower)
in expansions around the CP-conserving point
$\widetilde\varphi_\alpha=0^\circ$ as functions of $\tan\beta$
for the scenario~(\ref{eq:cpsps1a}).
The black, red, and blue solid lines are for the components of
$\Phi_1$, $\Phi_2$, and $\Phi_3$, respectively, and
the black, red, and blue dashed lines for the components of
$\Phi_{A_u}$, $\Phi_{A_d}$, and $\Phi_{A_e}$, respectively.
}
\label{fig:coeff.0}
\end{figure}
\begin{figure}[!t]
\begin{center}
{\epsfig{figure=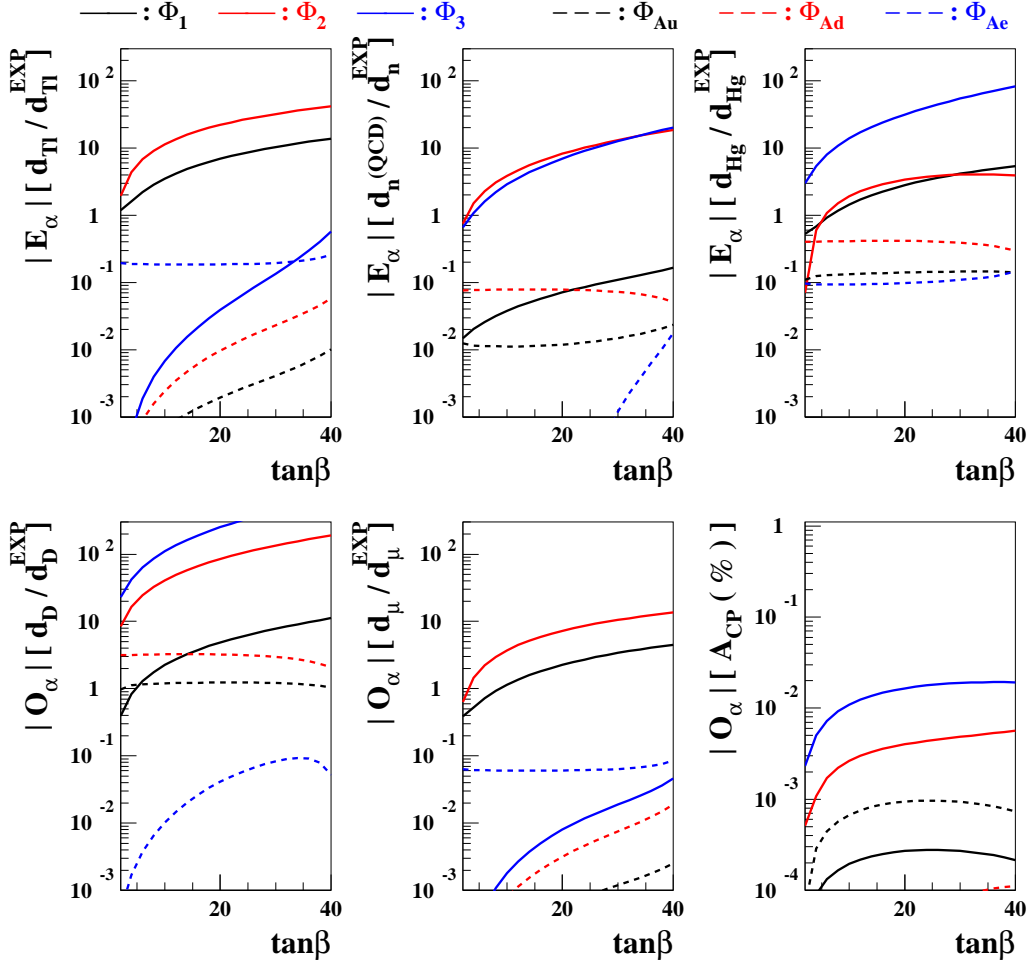,height=14.0cm,width=14.0cm}}
\end{center}
\vspace{-1.0cm}
\caption{\it The same as in Fig.~\ref{fig:coeff.0} but for expansions around
the CP-conserving point $\widetilde\varphi_\alpha=180^\circ$.
}
\label{fig:coeff.180}
\end{figure}
In Figs.~\ref{fig:coeff.0} and \ref{fig:coeff.180}, we show
the absolute values of the components of the 
six 6D vectors 
\begin{eqnarray}
&&
{\bf E}^{\,d_{\rm Tl}}\equiv\nabla (d_{\rm Tl}/d_{\rm Tl}^{\rm EXP})\,, \ \ \
{\bf E}^{\,d_{\rm n}}\equiv\nabla (d_{\rm n}/d_{\rm n}^{\rm EXP})\,, \ \ \
{\bf E}^{\,d_{\rm Hg}}\equiv\nabla (d_{\rm Hg}/d_{\rm Hg}^{\rm EXP})\,;
\nonumber \\
&&
{\bf O}^{\,d_{\rm D}}\equiv\nabla (d_{\rm D}/d_{\rm D}^{\rm EXP})\,, \ \ \
{\bf O}^{\,d_\mu} \equiv\nabla (d_{\mu}/d_{\mu}^{\rm EXP})\,, \ \ \
{\bf O}^{\,A_{\rm CP}}\equiv\nabla (A_{\rm CP}/\%)\,, 
\end{eqnarray}
around the CP-conserving points
$\widetilde\varphi_\alpha=0^\circ$ and $\widetilde\varphi_\alpha=180^\circ$, 
respectively, varying $\tan\beta$.
We do not show ${\bf O}^{\,\phi_{B_s}}$, since we find that this observable
is too small to be detectable in the class of scenarios under consideration.
The solid lines are for the components of the CP-violating 
phases of the gaugino mass parameters ($\alpha=1,2,3$)
and the dashed lines for those of the trilinear $A$ parameters
($\alpha=4,5,6$).
The dominant components are
$\left({\bf E}^{\,d_{\rm Tl}}\right)_{2,1}$, 
$\left({\bf E}^{\,d_{\rm n}}\right)_{2,3}$, 
$\left({\bf E}^{\,d_{\rm Hg}}\right)_{3,2}$, 
$\left({\bf O}^{\,d_{\rm D}}\right)_{3,2}$, 
$\left({\bf O}^{\,d_{\mu}}\right)_{2,1}$, and 
$\left({\bf O}^{\,A_{\rm CP}}\right)_{3,2}$, 
reflecting strong dependences on the CP-violating phases of the gaugino mass parameters. The
$\Phi_{1,2,3}$ components grow as $\tan\beta$ increases, implying that the
EDM constraints on the CP phases $\Phi_{1,2,3}$ become stronger for larger $\tan\beta$.
On the other hand, the $\Phi_{A_u,A_d,A_e}$ components are less than unity, except for
${\bf O}^{\,d_{\rm D}}$. This implies that the EDM constraints on these phases are weaker,
and that the Deuteron EDM may be large enough to be observed in the 
proposed experiment.
We also observe that the component $\left({\bf O}^{\,A_{\rm CP}}\right)_{A_u} > 10^{-2}$
when $\tan\beta \gsim 30$  and
$\widetilde\varphi_\alpha=0^\circ$, 
implying that it could give rise to 
the CP asymmetry larger than $1$ \% when $\Phi_{2,3}\sim 0^\circ$
but $\Phi_{A_u}$ is large, about $100^\circ$.

\begin{figure}[!t]
\begin{center}
{\epsfig{figure=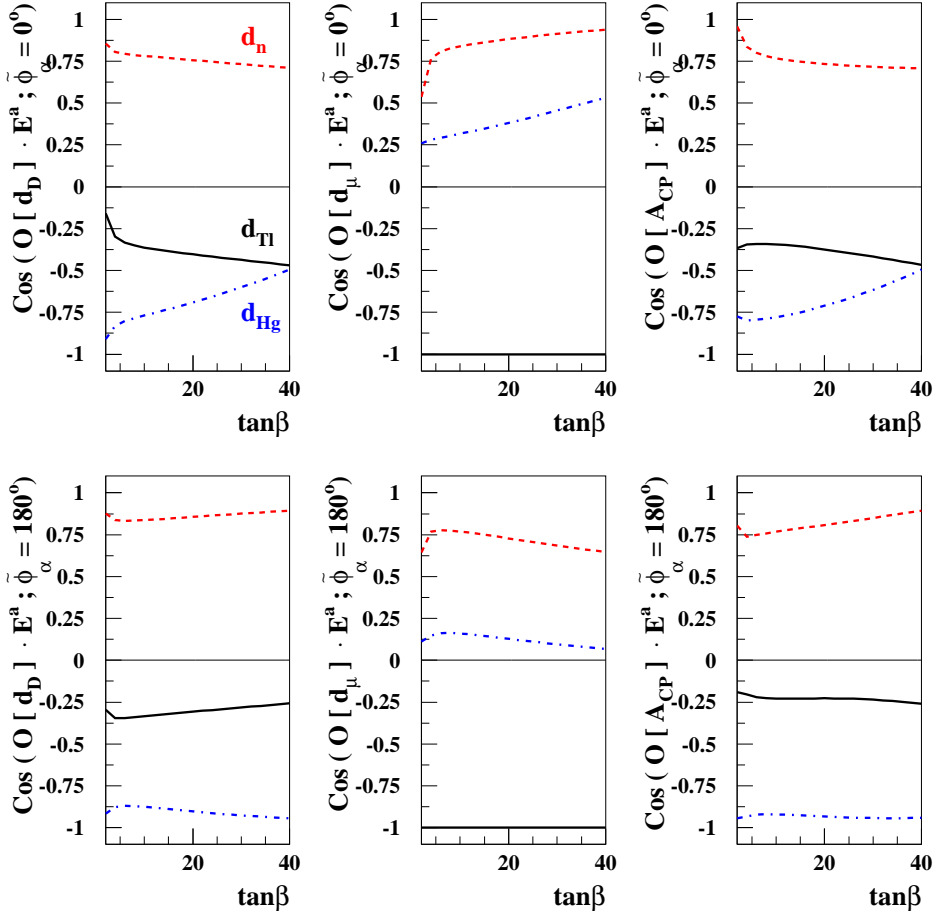,height=14.0cm,width=14.0cm}}
\end{center}
\vspace{-1.0cm}
\caption{\it The  cosines between the three observable
vectors ${\bf O}^{\,d_{\rm D}\,,d_\mu\,,A_{\rm CP}}$ (left, middle, right frames)
and the three EDM-constraint vectors
${\bf E}^{\,d_{\rm Tl}\,,d_{\rm n}\,,d_{\rm Hg}}$ (solid, dashed, dash-dotted lines)
as functions of $\tan\beta$
for $\widetilde\varphi_\alpha=0^\circ$ (upper)
and $\widetilde\varphi_\alpha=180^\circ$ (lower).
}
\label{fig:cosine}
\end{figure}
In Fig.~\ref{fig:cosine}, we display the cosines
$C_{{\bf  O,E}^a}$  of the  relative  angles  between the three observable
vectors ${\bf O}^{\,d_{\rm D}\,,d_\mu\,,A_{\rm CP}}$ 
and the three EDM-constraint vectors 
${\bf E}^{\,d_{\rm Tl}\,,d_{\rm n}\,,d_{\rm Hg}}$. The upper frames are for
the case of $\widetilde\varphi_\alpha=0^\circ$ and 
the lower ones for $\widetilde\varphi_\alpha=180^\circ$.
We see that the cosines between the observables and ${\bf E}^{\,d_{\rm Tl}}$ (black solid lines)
are reasonably small,
except the case of the muon EDM. In the case, the Thallium EDM is dominated
by the electron EDM, resulting in the high 
degeneracy between the Thallium EDM constraint and the muon EDM~\footnote{The two 
EDMs are not exactly degenerate, due to the
additional contribution to $d_{\rm Tl}$ from the electron-Nucleon
interaction $C_S\,\bar{e}i\gamma_5\,e\,\bar{N}N$, which
becomes larger as $\tan\beta$ increases. We find that the degeneracy is lifted by 
an amount of ${\cal O}(10^{-4})$ for large $\tan\beta$, which 
it is too small to have visible effect in the figure.}.
This alignment between the Thallium and muon EDMs in the scenario
under consideration
leads to a prediction of the muon EDM that is below the
projected sensitivity, as we show in the next Section.
On the other hand, the cosines between the ${\bf O}^{\,d_{\rm D}\,,A_{\rm CP}}$
and ${\bf E}^{\,d_{\rm n}}$ (red dashed lines)
are larger than those between ${\bf O}^{\,d_{\rm D}\,,A_{\rm CP}}$
and ${\bf E}^{\,d_{\rm Hg}}$ (blue dash-dotted lines) for
$\widetilde\varphi_\alpha=0^\circ$.
In the case $\widetilde\varphi_\alpha=180^\circ$, the cosines between the observable and the
Mercury EDM-constraint vectors are slightly larger.

Having the vectors representing the EDM constraints and observables in
hand, one can combine them  to construct the optimal directions in the
6D        space       of       CP-violating        phases,       using
Eq.~(\ref{Phi6D}).  Specifically,   we  consider  the   three  optimal
directions  that  maximize $d_{\rm  D}$,  $d_\mu$,  and $A_{\rm  CP}$,
respectively, taking into account  the three existing EDM constraints.
As  comparisons  to  them,   we  also  consider  two  other  reference
directions,    which   have    $\Delta\Phi_1=\Delta\Phi_{A_e}=0$   and
$\Delta\Phi_2=\Delta\Phi_3=0$,   where    $\Delta\Phi$   denotes   the
difference  from  the  corresponding  CP-conserving point.  These  two
reference directions can be constructed by defining
\begin{equation}
\Phi^*_\alpha\ \equiv \ {\cal N}\,
\varepsilon_{\alpha\beta\gamma\delta\mu\nu}\,
 E^a_\beta\, E^b_\gamma\, E^c_\delta\; N^{(1)}_\mu\, N^{(2)}_\nu\ ,
\end{equation}
where, for each direction, the two null directions $N^{(1,2)}_\mu$ are
chosen as
\begin{eqnarray}
N^{(1)}_\mu=(1,0,0,0,0,0)\,, N^{(2)}_\mu=(0,0,0,0,0,1)\, & {\rm
 for~the~direction} & 
 \Delta\Phi_1=\Delta\Phi_{A_e}=0 ,
\nonumber \\ 
N^{(1)}_\mu=(0,1,0,0,0,0)\,, N^{(2)}_\mu=(0,0,1,0,0,0)\, & {\rm
 for~the~direction} & 
\Delta\Phi_2=\Delta\Phi_3=0\ .
\nonumber \\ 
\end{eqnarray}

\begin{figure}[!t]
\begin{center}
{\epsfig{figure=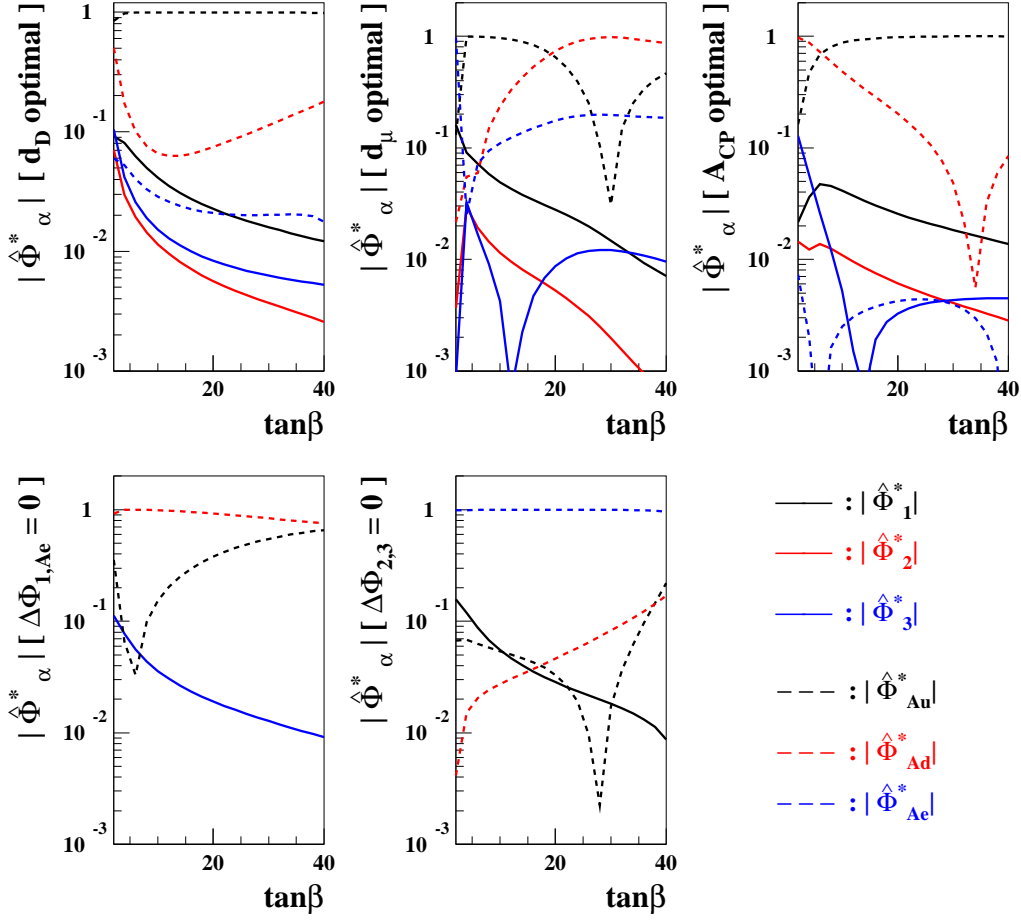,height=14.0cm,width=14.0cm}}
\end{center}
\vspace{-1.0cm}
\caption{\it The absolute values of the
six components of the five normalized direction vectors considered in the text
for $\widetilde\varphi_\alpha=0^\circ$. The solid lines represent the 
$\Phi_{1,2,3}$ components, and the dashed lines
the $\Phi_{A_u,A_d,A_e}$ components.
}
\label{fig:efd.0}
\end{figure}
\begin{figure}[!t]
\begin{center}
{\epsfig{figure=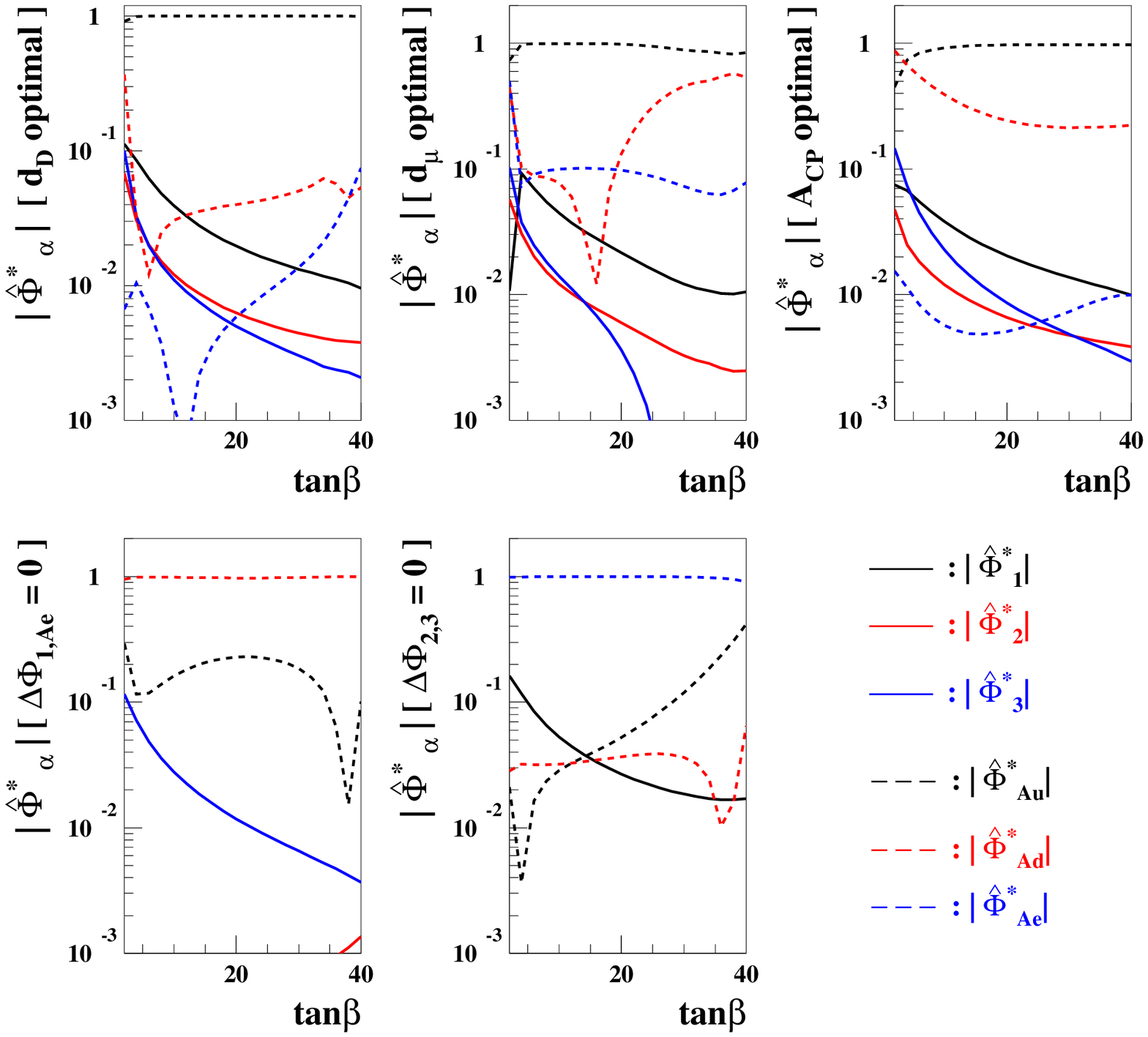,height=14.0cm,width=14.0cm}}
\end{center}
\vspace{-1.0cm}
\caption{\it The same as in Fig.~\ref{fig:efd.0}, but for
$\widetilde\varphi_\alpha=180^\circ$.
}
\label{fig:efd.180}
\end{figure}
We  show in Figs.~\ref{fig:efd.0}  and \ref{fig:efd.180}  the absolute
values  of the  six  components  of the  five  normalized vectors  for
$\widetilde\varphi_\alpha=0^\circ$                                  and
$\widetilde\varphi_\alpha=180^\circ$,  respectively. We  first observe
that the $\Phi_{1,2,3}$ components (solid lines) are relatively small,
and decrease  as $\tan\beta$ increases. Hence, all  the directions are
mostly given  by some combination of $\Phi_{A_u}$  (black dashed line)
and $\Phi_{A_d}$  (red dashed line) directions, with  the exception of
the $\Delta\Phi_{2,3}=0$  direction, which is mainly  aligned with the
$\Phi_{A_e}$ (blue dashed line) direction.  The $\Phi_{A_u}$ component
is  generally larger than  the $\Phi_{A_d}$  component in  the optimal
directions,     except    in    the     case    of     $d_\mu$    with
$\widetilde\varphi_\alpha=0^\circ$ and  $\tan\beta \gsim 20$,  as seen
in the middle-upper frame of Fig.~\ref{fig:efd.0}.

\begin{figure}[!t]
\begin{center}
{\epsfig{figure=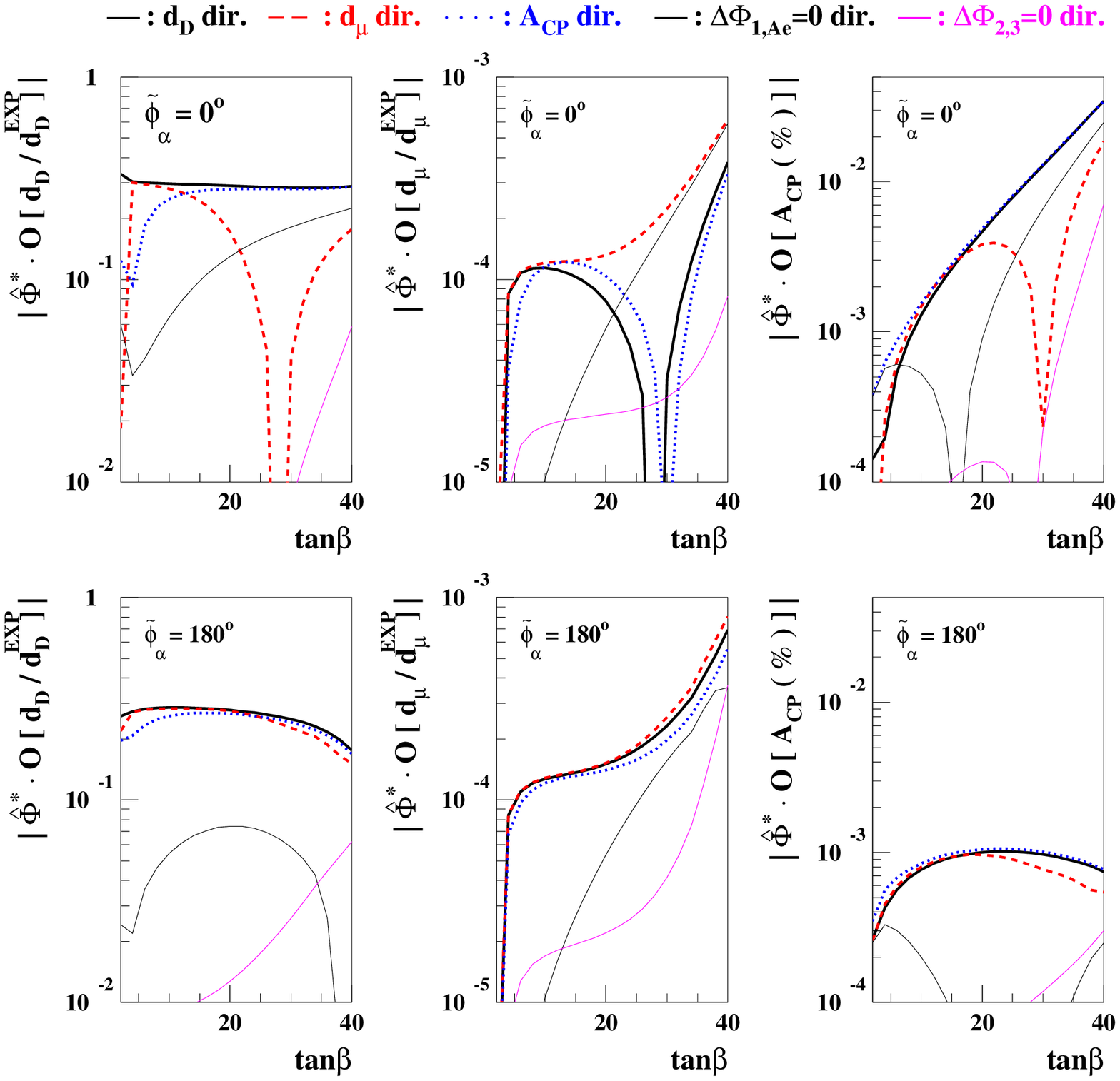,height=14.0cm,width=14.0cm}}
\end{center}
\vspace{-1.0cm}
\caption{\it The products $\widehat{\Phi}^* \cdot {\bf O}$
along the directions optimized for $d_{\rm D}$, $d_\mu$ and
$A_{\rm CP}$, which are denoted by the thick black solid, red
dashed and blue dotted lines, respectively. The thin lines are for the products
along the $\Delta\Phi_{1,A_e}=0$ (thin black)
and $\Delta\Phi_{2,3}=0$ (thin magenta) directions.
The upper frames are for
$\widetilde\varphi_\alpha=0^\circ$ and the lower ones for
$\widetilde\varphi_\alpha=180^\circ$.
}
\label{fig:coefd}
\end{figure}
Finally, we consider the  products $\widehat{\Phi}^* \cdot {\bf O}$ of
the  6D   vectors  of  the  normalized  optimal   directions  and  the
observables. The products determine the sizes of the observables along
the  directions through the  relations given  in Eq.~(\ref{Ooptimal6})
when $\phi^*=1$. As  we see in the next Section,  $\phi^*$ could be as
large as $\sim 100$  before the small-phase approximation breaks
down and  one of the  three EDM constraints  is violated.  We  show in
Fig.~\ref{fig:coefd}  the products  for the  directions  optimized for
$d_{\rm D}$, $d_\mu$ and $A_{\rm  CP}$, which are denoted by the thick
black solid, red dashed, and blue dotted lines, respectively. The thin
lines are for the  products along the reference $\Delta\Phi_{1,A_e}=0$
(thin black) and  $\Delta\Phi_{2,3}=0$ (thin magenta) directions.  The
upper frames are  for $\widetilde\varphi_\alpha=0^\circ$ and the lower
ones  for $\widetilde\varphi_\alpha=180^\circ$.   We observe  that the
directions constructed  using the geometric prescription  given in the
previous Section do indeed give the optimal values of the observables,
which  are larger  than those  along  the other  optimal and  reference
directions, sometimes even much larger, depending on $\tan\beta$.

\setcounter{equation}{0}
\section{Optimal Values of CP-Violating Observables in the 
MCPMFV SUSY Model}\label{sec:optimal}
%
As preparation  for presenting our numerical results  for $d_{\rm D}$,
$d_\mu$,  and $A_{\rm  CP}$, we  first examine  the magnitudes  of the
Thallium, neutron  and Mercury  EDMs along the  three optimal  and two
reference directions  considered in the previous Section.   As long as
the small-phase approximation is valid,  the EDMs should lie below the
current  experimental limits,  but the  approximation breaks  down for
large  phases,   leading  to   non-vanishing  EDMs  larger   than  the
limits. The breakdown of the small-phase approximation would limit the
maximum  values   of  $\phi^*$  in   Eq.~(\ref{Ooptimal6})  to  values
depending  on the  choice  of direction  and,  accordingly, limit  the
maximum values  of the  CP-odd observables that  can be found  in this
approach.

\begin{figure}[!t]
\begin{center}
{\epsfig{figure=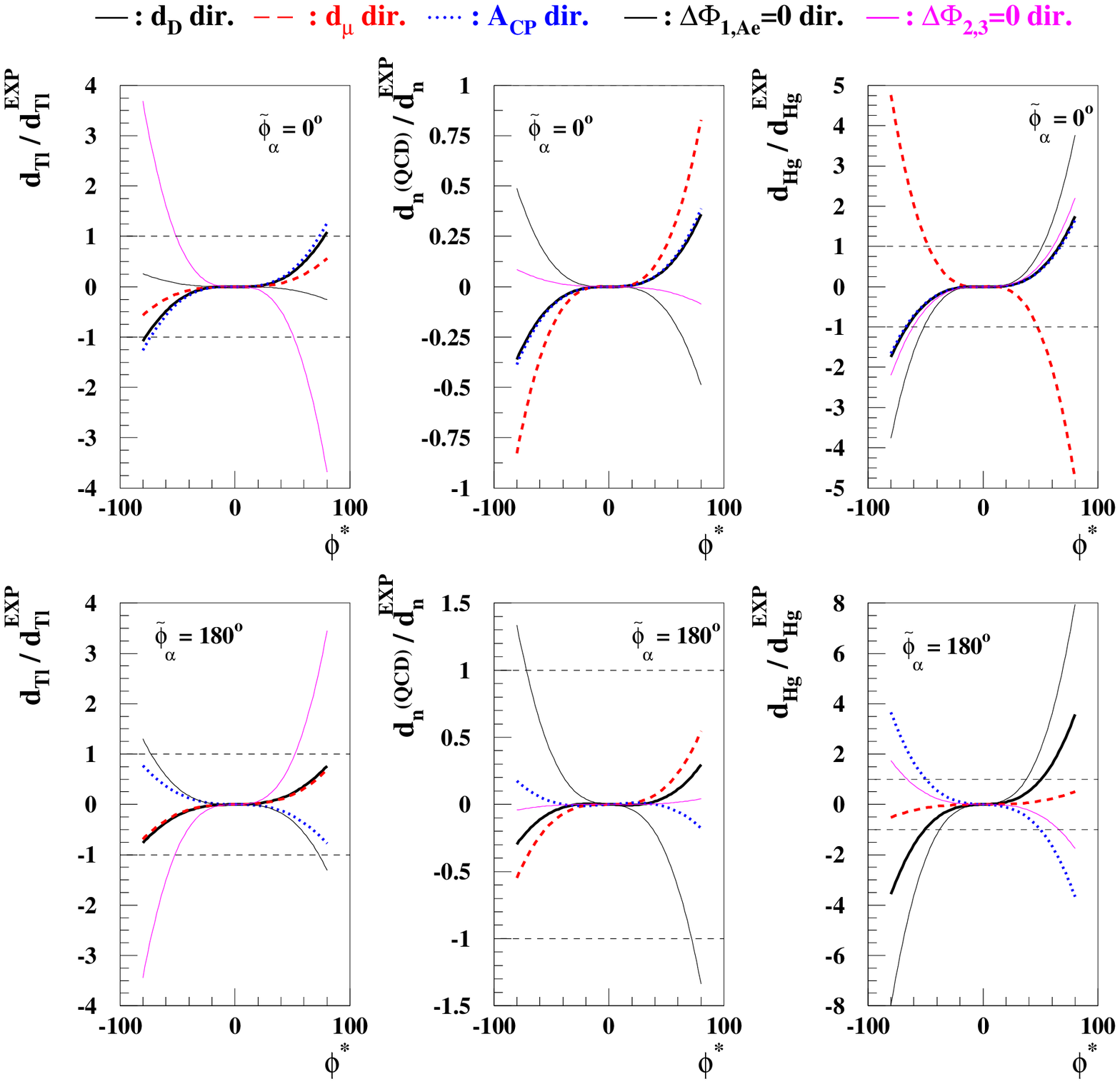,height=14.0cm,width=14.0cm}}
\end{center}
\vspace{-1.0cm}
\caption{\it  The values of the three EDMs along the three optimal
and two reference directions for
$\tan\beta=40$, with $\widetilde\varphi=0^\circ$ (upper)
and $\widetilde\varphi=180^\circ$ (upper). The line styles are the same as in
Fig.~\ref{fig:coefd}, with the additional horizontal lines representing the EDM constraints.
}
\label{fig:dtlnhg.40}
\end{figure}
\begin{figure}[!t]
\begin{center}
{\epsfig{figure=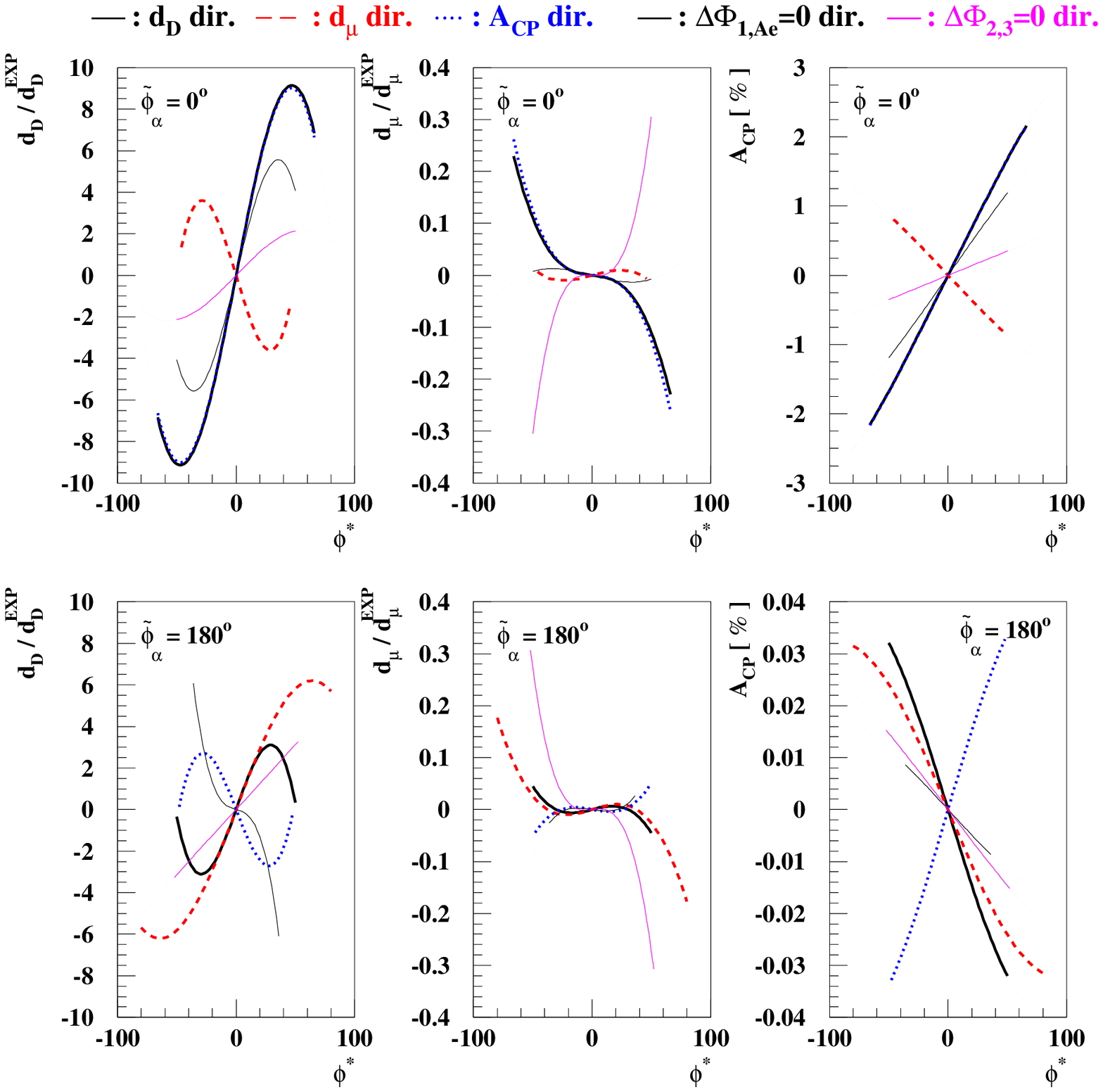,height=14.0cm,width=14.0cm}}
\end{center}
\vspace{-1.0cm}
\caption{\it  The three CP-odd observables
along the three optimal and two reference directions for
$\tan\beta=40$, with $\widetilde\varphi=0^\circ$ (upper)
and $\widetilde\varphi=180^\circ$ (upper). The line styles are the same as in
Fig.~\ref{fig:coefd}.
}
\label{fig:obs.40}
\end{figure}

Figure  \ref{fig:dtlnhg.40}  shows  the  three  constrained  EDMs  for
$\tan\beta=40$. From this Figure, one can read off the maximum allowed
value of  $\phi^*$ for  each direction.  For  example, from  the upper
frames   with  $\widetilde\varphi=0^\circ$,  $\left(\phi^*\right)^{\rm
  max}  \sim  70\,,  50\,,  70\,,  50\,,$ and  $50$  for  the  $d_{\rm
  D}$-optimal,      $d_{\mu}$-optimal,      $A_{\rm      CP}$-optimal,
$\Delta\Phi_{1,A_e}=0$,     and    $\Delta\Phi_{2,3}=0$    directions,
respectively, with the most important constraints being those provided
by $d_{\rm Hg}$, $d_{\rm Hg}$, $d_{\rm Hg}$, $d_{\rm Hg}$, and $d_{\rm
  Tl}$, respectively.   Inserting the values $\left(\phi^*\right)^{\rm
  max}$ into  the corresponding products  $\widehat{\Phi}^* \cdot {\bf
  O}$, one may get (near-)maximal values of the CP-odd observables in
the linear regime.

Fig.~\ref{fig:obs.40}  shows the  three CP-odd  observables  along the
five directions when $\tan\beta=40$.  The experimental upper bounds on
the Thallium, neutron and Mercury EDMs have been imposed.  In the case
of  the  Deuteron  EDM  with  $\widetilde\varphi_\alpha=0^\circ$,  the
linear regime ends  at $\phi^* \sim \pm 30$, and  beyond this range it
reaches  saturation at  a  value $\sim  |9|$, subsequently  decreasing
until  the  maximal   value  of  the  angle  $\left(\phi^*\right)^{\rm
max}_{\widetilde\varphi_\alpha=0^\circ}\sim  |70|$  is reached.   When
$\widetilde\varphi_\alpha=180^\circ$,   the  linear  regime   ends  at
$\phi^* \sim  \pm 30$. Beyond this  point, the Deuteron  EDM along the
$d_{\rm D}$ optimal direction  saturates, but it continues to increase
up  to  $\sim  |6|$   along  the  $d_\mu$  and  $\Delta\Phi_{1,A_e}=0$
directions.  {\it These two examples demonstrate that the existing EDM
constraints do not exclude the observation of $d_{\rm D}$.}

On the other hand, the muon EDM is always below the projected
sensitivity, since $\widehat{\Phi}^* \cdot {\bf O}^{\,d_\mu}$ is
forced to be small by the (near-) degeneracy with the Thallium EDM,
see Fig.~\ref{fig:coefd}.

\begin{figure}[!t]
\begin{center}
{\epsfig{figure=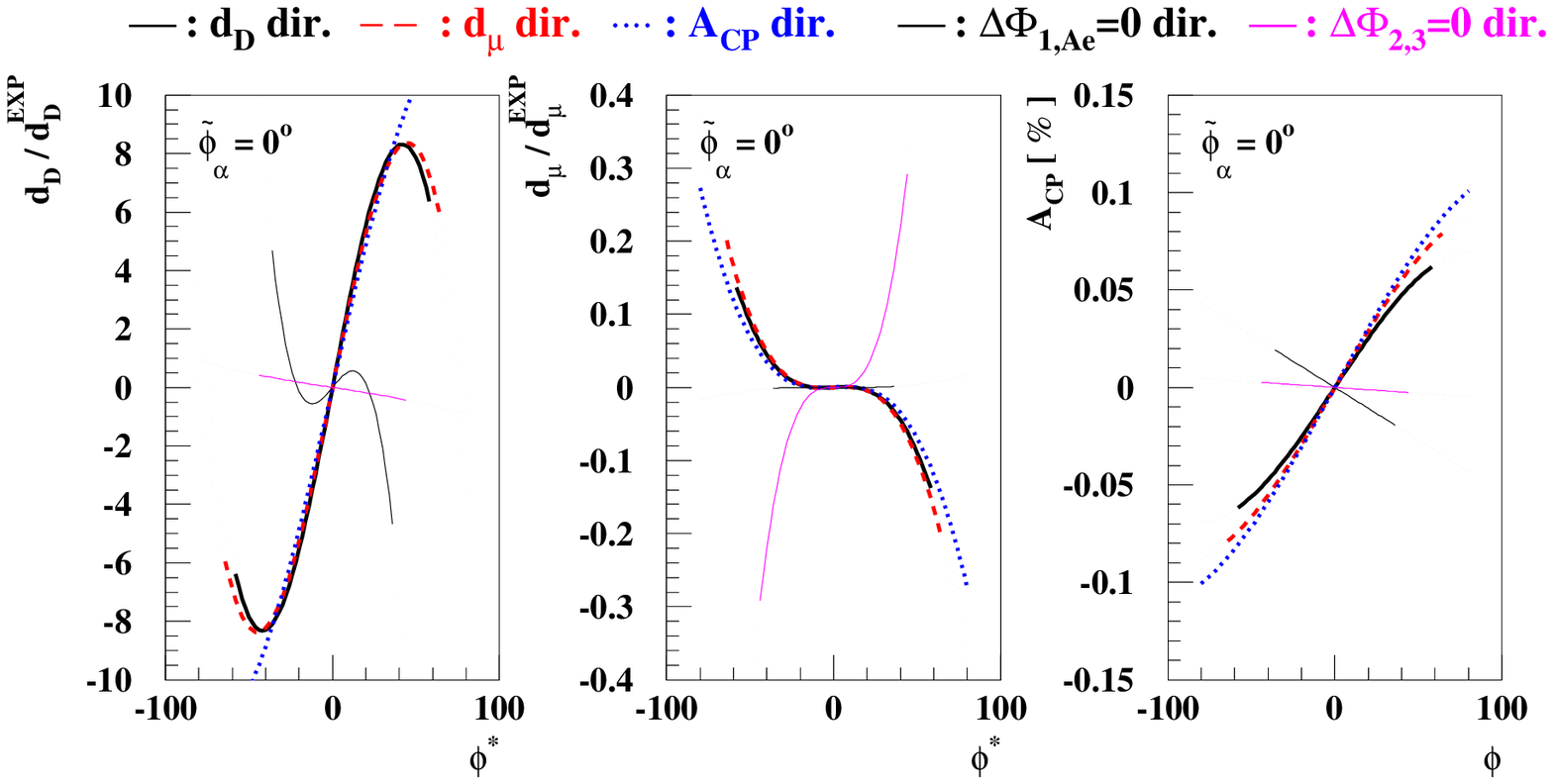,height=14.0cm,width=14.0cm}}
\end{center}
\vspace{-7.0cm}
\caption{\it  The same as in Fig.~\ref{fig:obs.40}, but for
$\tan\beta=10$ and $\widetilde\varphi=0^\circ$.
}
\label{fig:obs.10}
\end{figure}

The CP asymmetry $A_{\rm CP}(b\to s\gamma)$ has a larger linear regime,
extending over the almost whole range with $|\phi^*| \lsim 100$,
and could as large as $\sim 2 \%$ when $\widetilde\varphi=0^\circ$.
Such a value could be
observed at a future Super B Factory~\cite{Aushev:2010bq}. 
Unfortunately, 
the case with $\tan\beta \sim 40$ gives too small a branching ratio
$B(b\to s\gamma) \sim 1 \times 10^{-4}$~\footnote{See also~\cite{Ellis:2007kb}, 
in which the same scenario has
been analyzed.}.
In order to respect the constraint from $B(b\to s\gamma)$, we consider the case of
small $\tan\beta=10$ with $\widetilde\varphi=0^\circ$, see Fig.~\ref{fig:obs.10}. 
This case is compatible with
the $B(b\to s\gamma)$ constraint
at the 2-$\sigma$ level, but the predicted $A_{\rm CP}$ is too small to be
observed. On the other hand, the attainable values of $d_{\rm D}$ and $d_\mu$
are similar to those in the case when $\tan\beta=40$.

\begin{figure}[!t]
\begin{center}
{\epsfig{figure=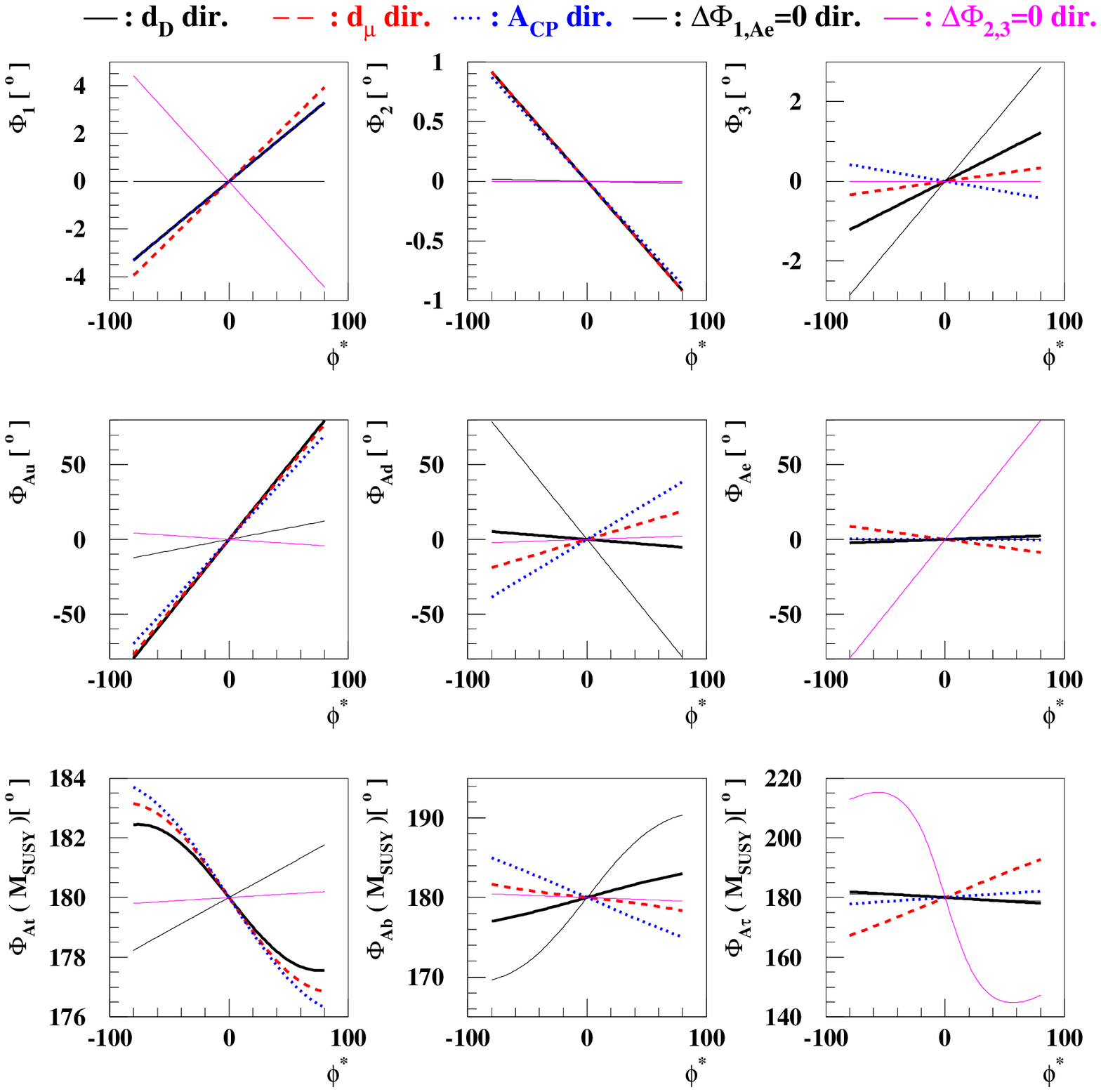,height=14.0cm,width=14.0cm}}
\end{center}
\vspace{-1.0cm}
\caption{\it  The 6 CP-violating phases at the GUT scale and the 3 CP-violating
phases of the third-generation
$A$ parameters at the SUSY scale along the 3 EDM-free directions
and 2 reference directions when
$\widetilde\varphi_\alpha=0^\circ$ and $\tan\beta=10$.
}
\label{fig:phi0.10}
\end{figure}

Finally, we  show in Fig.~\ref{fig:phi0.10} the  6 CP-violating phases
at the GUT scale and the 3 CP-violating phases of the third-generation
$A$       parameters      at       the      SUSY       scale      when
$\widetilde\varphi_\alpha=0^\circ$  and  $\tan\beta=10$.   We  observe
that  the CP-odd gaugino  phases $\Phi_1$,  $\Phi_2$ and  $\Phi_3$ can
only be as large as $4^\circ$, $1^\circ$ and $2^\circ$, respectively,
whereas  the CP-odd trilinear $A$-phases $\Phi_{A_u,A_d,A_e}$  at the
GUT scale can be as large as $\pm 70^\circ$. These CP-violating phases
are  suppressed  at  the  SUSY  scale  by  RG  running  from  the  GUT
scale~\cite{Ellis:2007kb},   but  sizeable   non-trivial  CP-violating
phases are still allowed at the SUSY scale: $\Delta\Phi_{A_t} \sim \pm
4^\circ$,     $\Delta\Phi_{A_b}     \sim     \pm    10^\circ$,     and
$\Delta\Phi_{A_\tau}  \sim \pm 40^\circ$.   Similar magnitudes  of the
CP-violating         phases         are         attainable         for
$\widetilde\varphi_\alpha=180^\circ$.

\setcounter{equation}{0}
\section{The 7D Extension Including $\theta_{\rm QCD}$}\label{sec:thetaQCD}
%
In the previous Sections, we implicitly assumed that the CP-violating QCD
$\theta$-term: 
\begin{equation}
{\cal L} = \frac{\alpha_s}{8\pi}\,\bar\theta\,G^a_{\mu\nu}\tilde{G}^{\mu\nu ,a}
\label{theta}
\end{equation}
vanishes,                   where                   $\tilde{G}^{\mu\nu
,a}=\epsilon^{\mu\nu\rho\sigma}G^a_{\rho\sigma}/2$  and  the parameter
$\bar\theta$ is given by the sum of the QCD $\theta_{\rm QCD}$ and the
strong chiral phase for the quark mass matrix as
\begin{equation} 
\bar\theta = \theta_{\rm QCD} +{\rm Arg}\,{\rm Det}\,{M_q}\,.
\end{equation} 
In the  weak basis  where ${\rm Arg}\,{\rm  Det}\,{M_q}$ = 0,  we have
$\bar\theta=\theta_{\rm  QCD}$.  The  QCD  $\theta$-term (\ref{theta})
would be  set to zero, e.g.,  if there is  a QCD axion in  the theory.
Otherwise, the dimension-four  operator (\ref{theta}) would in general
contribute to  the neutron, Mercury  and Deuteron EDMs,  e.g., through
the CP-odd pion-nucleon-nucleon interactions
\begin{equation}
{\cal L}_{\pi NN}=\bar{g}^{(0)}_{\pi NN}\,\overline{N}\tau^a N\,\pi^a \ + \
\bar{g}^{(1)}_{\pi NN}\,\overline{N}N\,\pi^0\,.
\end{equation}
QCD sum rule techniques have been used to estimate
the contribution of the $\bar\theta$ term
to the neutron EDM in~\cite{Pospelov:2000bw,Pospelov:2005pr}~\footnote{Though 
there are some inconsistencies
between the two references,
the second numerical equation in~\cite{Pospelov:2005pr}
is not affected by them.}:
\begin{eqnarray}
d_n(\bar\theta) &=& (0.4\pm 0.2) \left[\chi (4e_d-e_u) m_* \,
\bar\theta \,
\right]
\nonumber \\ &=&
(1\pm0.5) \frac{|\langle\bar{q}q\rangle |}{(225~{\rm MeV})^3}\, 
\bar\theta \,
\times 2.5\times 10^{-16}\,e\,{\rm cm} ,
\end{eqnarray}
where the
reduced mass $m_*=m_um_d/(m_u+m_d)$, $e_d=-(1/3)\,e$, $e_u=(2/3)\,e$, and
the condensate susceptibility $\chi=-5.7\pm0.6~{\rm GeV}^{-2}$
\footnote{  It has  been argued  that, in  the presence  of  an axion,
$\bar\theta$  should  be  replaced  by $\theta_{\rm  ind}$,  which  is
determined dynamically  via the  chromoelectric dipole moments  of the
up,  down, and  strange quarks  $d^C_{u,d,s}$.  However, since we lack full
knowledge       of      all       the       relevant      higher-order
corrections~\cite{private:PospelovRitz},
we  will  neglect $\theta_{\rm  ind}$  contributions  to  EDMs in  our
analysis.}.  The $d_n(\bar\theta)$ contribution  is to be added to the
other EDM  contributions from  the electric and  chromoelectric dipole
moments  of  light quarks,  the  Weinberg  operators,  and the  CP-odd
four-fermion interactions that are  induced by the CP-violating phases
in the soft SUSY-breaking sector,  i.e., by the six independent MCPMFV
CP phases considered above~\cite{ELPEDM}.
%

The leading contribution  of the $\bar\theta$ term
to the Mercury EDM is expected to be through its contribution to the
pion-nucleon-nucleon interaction coefficient $\bar{g}^{(1)}_{\pi NN}$:
\begin{equation}
d_{\rm Hg}(\bar\theta) = +(1.8\times 10^{-3}\,{\rm GeV}^{-1})
\,e\,\bar{g}^{(1)}_{\pi NN}(\bar\theta)\,.
\end{equation}
The  $\bar\theta$ contribution to the coupling is suppressed by
the factor $(m_d-m_u)/m_s$, and given by~\cite{Lebedev:2004va}
\begin{equation}
\bar{g}^{(1)}_{\pi NN}(\bar\theta)=
\frac{m_*\,\bar\theta}{f_\pi}\,\frac{m_d-m_u}{4m_s}
\langle N | \bar{u}u+\bar{d}d-2\bar{s}s| N\rangle
\simeq 1.1\times10^{-3}\,\bar\theta\ ,
\end{equation}
which results in
\begin{equation}
d_{\rm Hg}(\bar\theta) \simeq +2.0\times 10^{-6}\,\bar\theta\,e\,{\rm GeV}^{-1}
\simeq 3.9\times 10^{-20}\,\bar\theta\,e\,{\rm cm}\,.
\end{equation}
%
%
The leading contributions  of the $\bar\theta$ term
to the Deuteron EDM are given by~\cite{Lebedev:2004va}
\begin{eqnarray}
d_D(\bar\theta) \simeq
-e \left[(3.5\pm 1.4) + (1.4\pm 0.4)\right]\times 10^{-3}\,\bar\theta\,{\rm GeV}^{-1}\,
\simeq -9.7\times 10^{-17}\,\bar\theta\,e\,{\rm cm}\,.
\end{eqnarray}
The first term is the leading-order QCD sum-rule estimate
of the $\bar\theta$ contribution to the 
sum of the proton and neutron EDMs, which enters only via
subleading isospin-violating corrections.
The second term arises from the coupling $\bar{g}^{(1)}_{\pi NN}(\bar\theta)$:
\begin{equation}
d_{\rm D}^{\pi NN}(\bar\theta) = -\ \frac{e\, g_{\pi NN}\,
\bar{g}^{(1)}_{\pi NN}(\bar\theta)}{12\pi m_\pi}\, \frac{1+\xi}{(1+2\xi)^2}\
\simeq\ -\, (1.3\pm 0.3)\,e\, \bar{g}^{(1)}_{\pi NN}\,{\rm
  GeV}^{-1}\, ,
\end{equation}
where $g_{\pi NN}\simeq 13.45$ and $\xi=\sqrt{m_p\epsilon}/m_\pi$, with
$\epsilon =2.23$  MeV being  the Deuteron binding  energy.
%
%

Including non-vanishing $\bar\theta$ together with the
six MCPMFV CP phases $\Phi_{1,2,3}$ and $\Phi_{A_u,A_d,A_e}$, 
there is a total of seven CP phases. The six-dimensional
geometric construction of optimal EDM-free directions in Section~2
can easily be extended by one more dimension in this case.
As in the 6D case,
the 7D EDM-constraint and observable vectors are given by
${\bf E}=\nabla E$ and ${\bf O}=\nabla O$, but with
\begin{equation}
\nabla_\alpha    \equiv    \left(
\frac{\partial}{\partial\Phi_1}, \
\frac{\partial}{\partial\Phi_2},    \
\frac{\partial}{\partial\Phi_3}, \
\frac{\partial}{\partial\Phi_{A_u}},  \
\frac{\partial}{\partial\Phi_{A_d}}, \
\frac{\partial}{\partial\Phi_{A_e}}, \
\frac{\partial}{\partial\widehat\theta}\right)\,.
\end{equation}
As before the CP-violating phases $\Phi_{1,2,3}$ and $\Phi_{A_u,A_d,A_e}$
are specified in degrees and 
we normalize $\bar\theta$ in units of $10^{-10}$:
\begin{equation}
\widehat\theta \equiv \bar\theta \times 10^{10}\,.
\end{equation}
With this normalization, with $\widehat\theta = 1$,
we have $d_n(\bar\theta)=2.5\times 10^{-26}\, e\,{\rm cm}$ which
is very near to the current experimental bound
$d_n^{\rm EXP}=3 \times 10^{-26}\, e\,{\rm cm}$.

\begin{figure}[!t]
\begin{center}
{\epsfig{figure=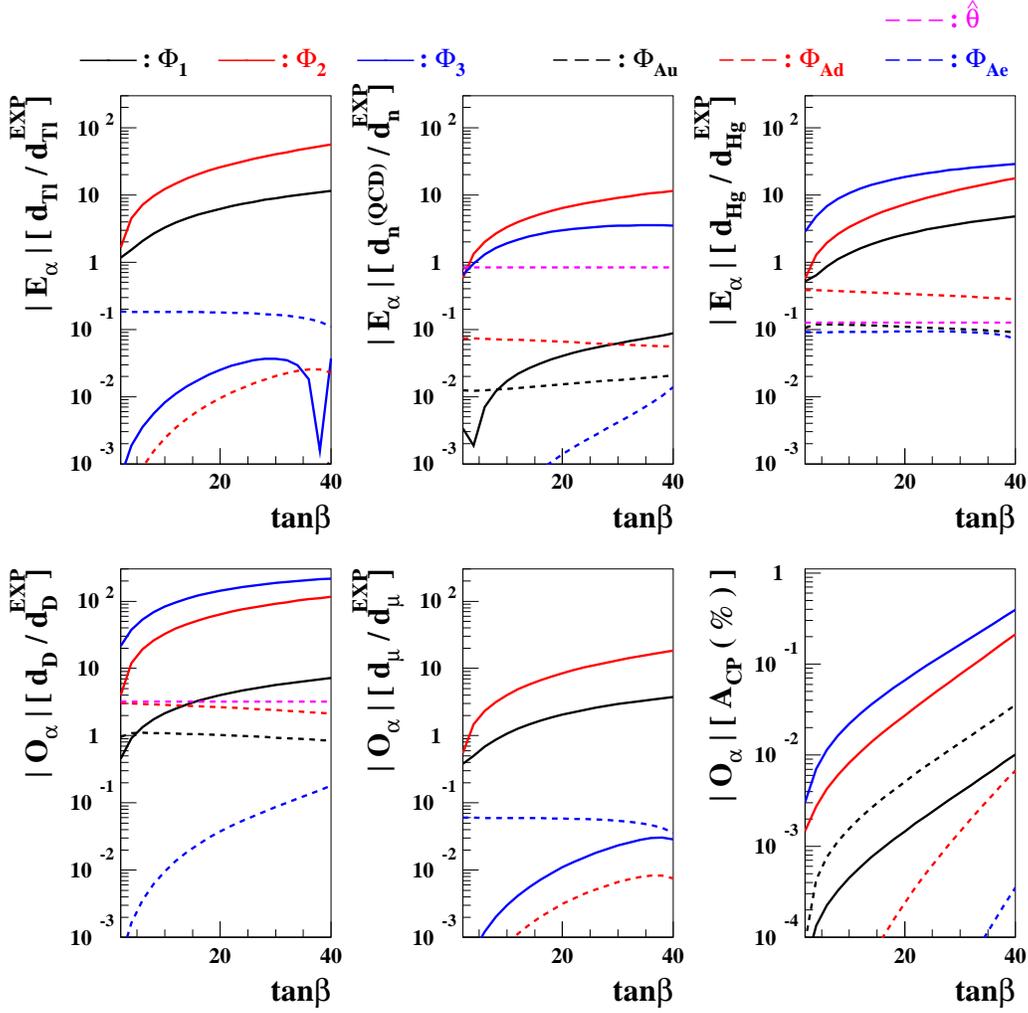,height=14.0cm,width=14.0cm}}
\end{center}
\vspace{-1.0cm}
\caption{\it The absolute values of the components of the three 7D EDM-constraint
vectors (upper) and those of the three 7D observable vectors (lower) in
linear expansions around $\widetilde\varphi_{\alpha=1-6}=0^\circ$, 
as functions of $\tan\beta$
for the scenario~(\ref{eq:cpsps1a}).
The magenta dashed lines represent the 
$\widehat\theta = \bar\theta \times 10^{10}$ component, while
the other lines are the same as in Fig.~\ref{fig:coeff.0}.
}
\label{fig:coeff.0.theta}
\end{figure}
In Fig.~\ref{fig:coeff.0.theta}, we show the absolute values of 
the components of the three 7D EDM-constraint
and the three 7D observable vectors in the linearized CP-violating
version of the scenario~(\ref{eq:cpsps1a}), taking 
$\widetilde\varphi_{\alpha=1-6}=0^\circ$.
Comparing with Fig.~\ref{fig:coeff.0} of the 6D case, we see that:
$(i)$ ${\bf E}^{d_{\rm Tl}}$,
${\bf O}^{d_\mu}$, and
${\bf O}^{A_{\rm CP}}$ are unchanged, with
vanishing seventh $\widehat\theta$ components,
$(ii)$ the first six components 
of ${\bf E}^{d_{\rm n}}$,
${\bf E}^{d_{\rm Hg}}$ and ${\bf O}^{d_{\rm D}}$ are
unchanged, and we have the new $\widehat\theta$ components,
$\left({\bf E}^{d_{\rm n}} \right)_7 \simeq 0.83$,
$\left({\bf E}^{d_{\rm Hg}} \right)_7 \simeq 0.13$ and
$\left({\bf E}^{d_{\rm D}} \right)_7 \simeq -3.2$,
which are independent of $\tan\beta$.

Having obtained the 7D ${\bf E}$ and ${\bf O}$ vectors, one can construct
the optimal  EDM-free  direction maximizing $O$ as 
in the 6D case:
\begin{equation}
  \label{Phi7D}
\Phi^*_\alpha\ =\ 
{\cal N}\; \varepsilon_{\alpha\beta\gamma\delta\mu\nu\rho}\,
 E^{d_{\rm Tl}}_\beta\, E^{d_{\rm n}}_\gamma\, E^{d_{\rm Hg}}_\delta\;
 B_{\mu\nu\rho}\; ,
\end{equation}
where the 3-form $B_{\mu\nu\rho}$ is given by
\begin{equation}
  \label{B3form}
B_{\mu\nu\rho}\ =\ \varepsilon_{\mu\nu\rho\lambda\sigma\tau\omega}\,
O_\lambda\, 
E^{d_{\rm Tl}}_\sigma\, E^{d_{\rm n}}_\tau\, E^{d_{\rm Hg}}_\omega\; .
\end{equation}
\begin{figure}[!t]
\begin{center}
{\epsfig{figure=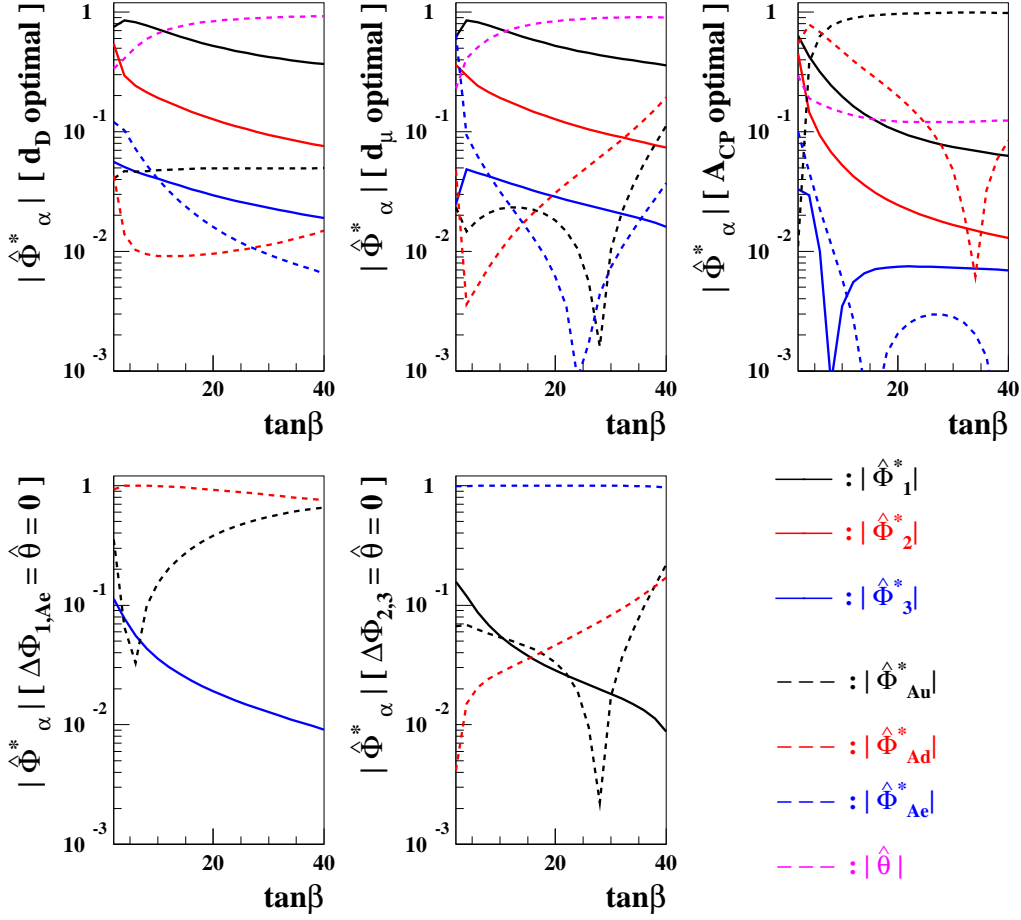,height=14.0cm,width=14.0cm}}
\end{center}
\vspace{-1.0cm}
\caption{\it The absolute values of the
seven components of the five normalized direction vectors 
for the 7D CP-violating version of the
scenario~(\ref{eq:cpsps1a}), in a linear expansion
around $\widetilde\varphi_{\alpha=1-6}=0^\circ$. 
The lines are the same as in Fig.~\ref{fig:efd.0}, with additional
magenta lines for
the $\widehat\theta$ components.
}
\label{fig:efd.0.theta}
\end{figure}

We show in Fig.~\ref{fig:efd.0.theta} the absolute values of the
seven components of the three optimum directions (upper) and
the two reference directions with 
$\Delta\Phi_1=\Delta\Phi_{A_e}=\widehat\theta=0$ (lower left)
and $\Delta\Phi_2=\Delta\Phi_3=\widehat\theta=0$ (lower middle),
considering the scenario~(\ref{eq:cpsps1a}) with  
$\widetilde\varphi_{\alpha=1-6}=0^\circ$.
Comparing with Fig.~\ref{fig:efd.0} of the 6D case, we see that:
$(i)$ the vectors in the two reference directions with $\widehat\theta=0$
remain the same,
$(ii)$ the $d_{\rm D}$- and $d_\mu$-optimal directions 
can now have sizeable $\Phi_{1,2}$ components, while
the $\widehat\theta$ component dominates when $\tan\beta \gsim 10$,
$(iii)$ the $A_{\rm CP}$-optimal direction is still dominated by
the $\Phi_{A_u}$ component when $\tan\beta \gsim 7$.
\begin{figure}[!t]
\begin{center}
{\epsfig{figure=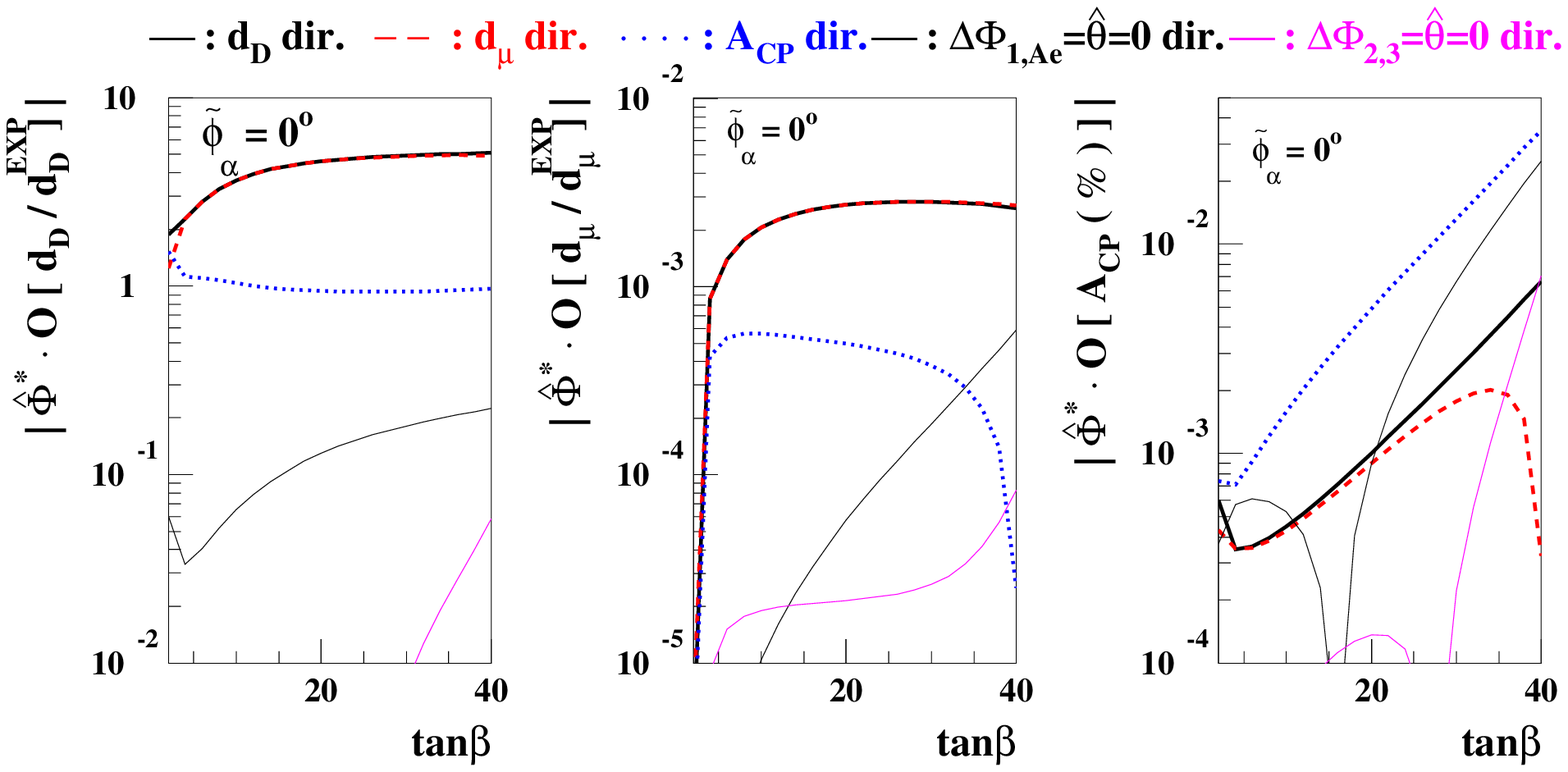,height=14.0cm,width=14.0cm}}
\end{center}
\vspace{-7.0cm}
\caption{\it The products $\widehat{\Phi}^* \cdot {\bf O}$
along the three $d_{\rm D}$-, $d_\mu$-, and
$A_{\rm CP}$-optimal and the two arbitrary directions 
for the 7D CP-violating version of the
scenario~(\ref{eq:cpsps1a}) with
$\widetilde\varphi_{\alpha=1-6}=0^\circ$.
The lines are the same as in Fig.~\ref{fig:coefd}.
}
\label{fig:coefd.0.theta}
\end{figure}
Fig.~\ref{fig:coefd.0.theta}  shows   the  products  with   the  three
observable vectors  of the three  optimal EDM-free directions  and the
two reference directions for the  same scenario. We observe again that
the optimal direction found using our geometric construction gives the
largest  value  for  each  corresponding observable.   Comparing  with
Fig.~\ref{fig:coefd} (upper  frames) of the 6D case,  the products can
be larger by more than an order of magnitude for the Deuteron and muon
EDMs. On the other hand, they remain more or less the same for $A_{\rm
CP}$,  due to  the  dominance  of the  $\Phi_{A_u}$  component in  the
$A_{\rm CP}$-optimal direction.

\begin{figure}[!t]
\begin{center}
{\epsfig{figure=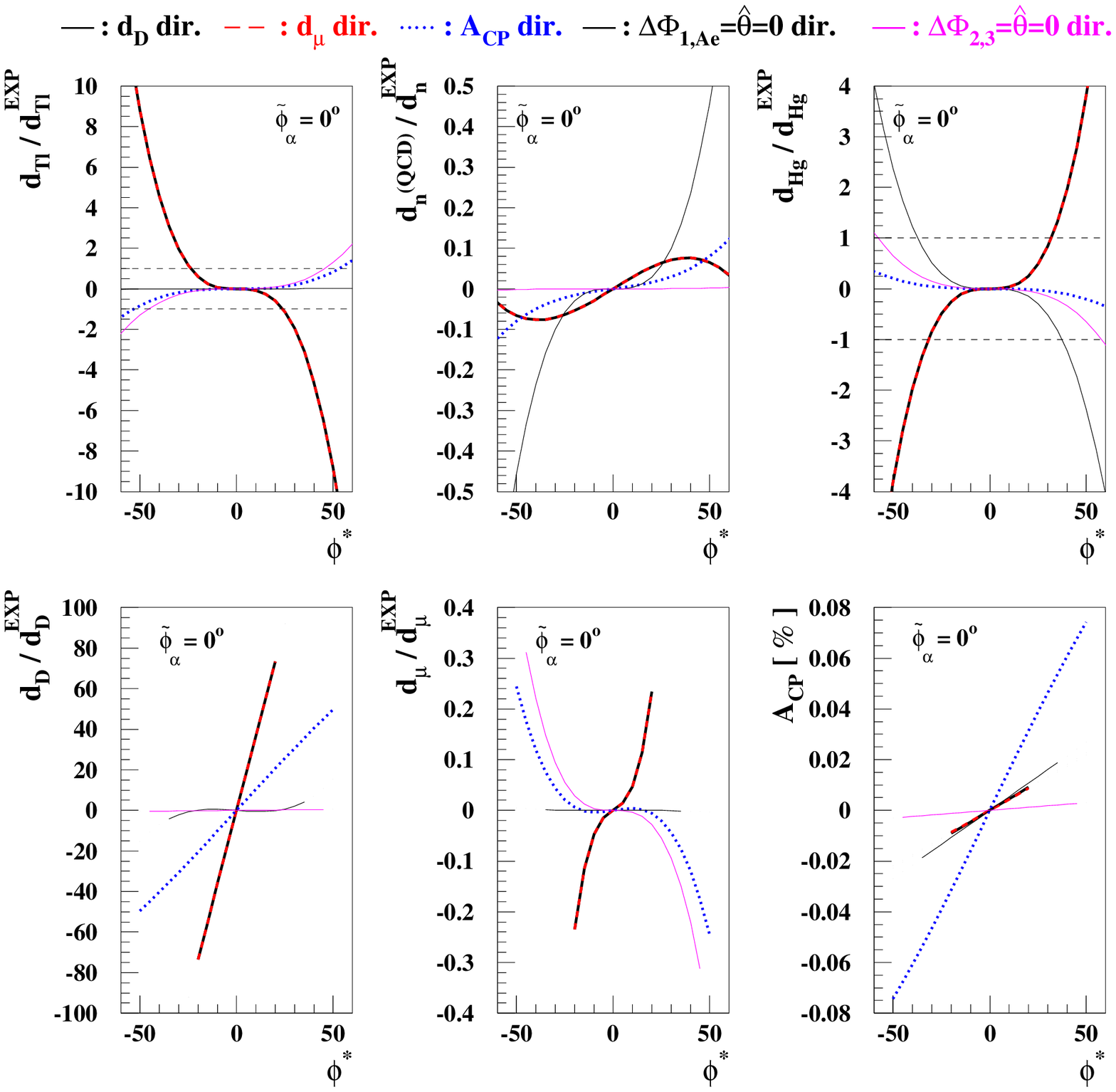,height=14.0cm,width=14.0cm}}
\end{center}
\vspace{-1.0cm}
\caption{\it 
Upper:  The three EDM constraints along the three EDM-free 
and two reference directions for
the scenario~(\ref{eq:cpsps1a}) with
$\tan\beta=10$ and $\widetilde\varphi_{\alpha=1-6}=0^\circ$.
Lower: The three CP-odd observables in the same scenario.
The lines are the same as in
Figs.~\ref{fig:dtlnhg.40} and \ref{fig:obs.40}.
}
\label{fig:edmobs.0.theta}
\end{figure}
In the upper frames of Fig. \ref{fig:edmobs.0.theta},
we show the three EDM constraints,
assuming $\tan\beta=10$ and $\widetilde\varphi_{\alpha=1-6}=0^\circ$.
As in the case of Fig.~\ref{fig:dtlnhg.40},
one can read off the
maximum values of $\phi^*$ for each EDM-free direction from the figure:
$\left(\phi^*\right)^{\rm max} \sim
25\,, 25\,, 50\,, 40\,,$ and $45$ for the
$d_{\rm D}$-optimal,
$d_{\mu}$-optimal,
$A_{\rm CP}$-optimal,
$\Delta\Phi_{1,A_e}=\widehat\theta=0$, and
$\Delta\Phi_{2,3}=\widehat\theta=0$ directions, which are mainly constrained by
$d_{\rm Tl}$,
$d_{\rm Tl}$,
$d_{\rm Tl}$,
$d_{\rm Hg}$, and
$d_{\rm Tl}$, respectively.
Multiplying the values $\left(\phi^*\right)^{\rm max}$ to the
corresponding products $\widehat{\Phi}^* \cdot {\bf O}$ shown in
Fig.~\ref{fig:coefd.0.theta},
we find the maximum values of the CP-odd observables in the linear regime,
which are shown in the lower frames of Fig. \ref{fig:edmobs.0.theta}.
Comparing with Fig.~\ref{fig:obs.10} of the 6D case, we see that the
maximal value of the Deuteron EDM is greatly enhanced, becoming as 
large as $\sim 70$ times the projected sensitivity.
On the other hand, the muon EDM is still below the projected sensitivity, and
the the maximal value of the CP asymmetry $A_{\rm CP}(b\to s\gamma)$ 
could be only $\sim 0.08$\%.

\begin{figure}[!t]
\begin{center}
{\epsfig{figure=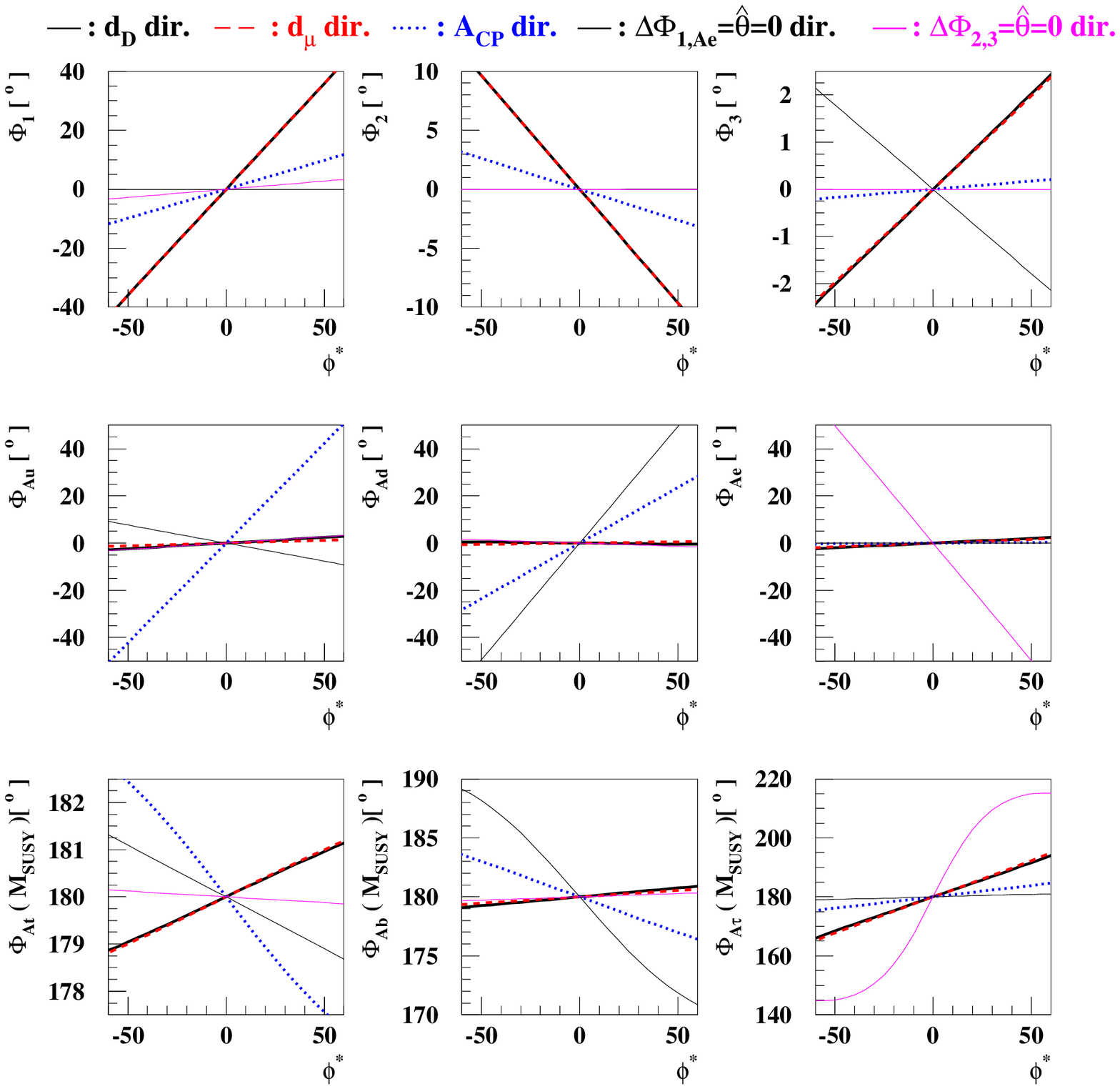,height=14.0cm,width=14.0cm}}
\end{center}
\vspace{-1.0cm}
\caption{\it  The 6 CP phases at the GUT scale and the 3 CP-violating 
phases of the third-generation
$A$ parameters at the SUSY scale along the 3 EDM-free and 2 reference directions.
The scenario~(\ref{eq:cpsps1a}) is taken with  $\tan\beta=10$ and
$\widetilde\varphi_{\alpha=1-6}=0^\circ$.
The line styles are the same as in Fig.~\ref{fig:phi0.10}.
}
\label{fig:phi0.10.theta}
\end{figure}
Finally, in  Fig.~\ref{fig:phi0.10.theta}, we show  the 6 CP-violating
phases at  the GUT scale and  the 3 CP phases  of the third-generation
$A$  parameters at  the SUSY  scale  along the  3 EDM-free  and the  2
reference    directions    in   the    7D    case.   Comparing    with
Fig.~\ref{fig:phi0.10} for the  6D case, we observe in  the top panels
that $\Phi_1$  and $\Phi_2$ can  be substantially larger, as  large as
$\sim    20^\circ$    and    $\sim   5^\circ$,    respectively,    for
$\left(\phi^*\right)^{\rm  max} \sim  25$ along  the $d_{\rm  D}$- and
$d_\mu$-optimal  directions  denoted by  the  thick  solid and  dashed
lines. We also note in the middle and bottom panels that the phases of
$A_{d,u,e}$ are somewhat larger than in the 6D case.  Finally, we note
(not  shown) that $\bar\theta$  could be  as large  as $\sim  5 \times
10^{-9}$ along the $A_{\rm CP}$-optimal direction.

\newpage
\section{Conclusions}

We  have  proposed in  this  paper  a  novel geometric  technique  for
optimizing  the possible  values  of CP-violating  observables in  the
presence of the  strong constraints due to upper  limits on EDMs.  Our
geometric approach  enables us to  separate the EDM-free  subspace off
the  full  CP-phase  parameter  space  in  the  linear  approximation.
Knowing the parametric  dependence of a given observable,  we can {\em
analytically} construct  the extremal direction in  the full parameter
space, along which the  observable gets maximized.  Since our approach
is analytic, it becomes exact  in the linear approximation and is much
more efficient and  accurate than the naive type of scan  that is usually made
in literature~\cite{RefSCAN}~\footnote{For comparison, we note that, for a theory with 6
free parameters, a search within  a 6-dimensional grid with
100 points in each coordinate would require $10^{12}$ scan  points.  Instead, our
geometric   method   only  involves   straightforward   sums  over   a
6-dimensional Levi-Civita tensor  with $6\,!=720$ non-zero components,
along a radial line of 50 points.}

We have demonstrated the applicability of this technique in two cases:
the 6D case of  the MCPMFV version of the MSSM, and  a 7D extension to
include the QCD vacuum phase. We have illustrated this approach within
a  class  of  CP-violating  models  that  extend  and  generalize  the
well-studied  SPS1a benchmark point  of the  CMSSM.  For  any specific
benchmark point, the values of CP-violating observables are in general
bounded  in magnitude,  since the  ranges of  the  CP-violating phases
$\phi_i$ are all  compact: $\phi_i \in [0, 2 \pi  )$.  Based on linear
expansions  of CP-violating  observables  around CP-conserving  points
with $\phi_i =  0, \pi$, our approach gives  rather accurate estimates
of the true maximal values of the CP-violating observables.

Using  this approach  in the  6D MCPMFV  case, we  find values  of the
Deuteron  EDM  that may  be  an order  of  magnitude  larger than  the
prospective  experimental  sensitivity.  This  range  is increased  by
almost   another  order   of  magnitude   if  our   optimal  geometric
construction is extended to the 7D case that includes the CP-violating
QCD vacuum phase. Hence, the  Deuteron EDM may provide indirect useful
information about the possible  presence of a non-vanishing QCD vacuum
phase,  complementing   the  current  experiments  on   EDMs  and  the
experimental  searches for  axions~\cite{KZ}.  On  the other  hand, we
find that  the maximal values of  the muon EDM are  somewhat below the
likely experimental sensitivity in both the scenarios with and without
the QCD  phase.  We also find  that the CP-violating $b  \to s \gamma$
decay asymmetry  $A_{\rm CP}$  is too small  to be observed,  once the
stringent  constraint from  $B(b\to s\gamma)$  is taken  into account.
Likewise, the $B_s$-mixing phase $\phi_{B_s}$ turns out to be close to
the small SM value in both the scenarios studied.

Our   geometric   approach  could   easily   be   extended  to   other
supersymmetric scenarios.  For example,  we have not made a systematic
survey of all the possibilities in the MCPMFV SUSY model that arise as
generalizations  of other  CP-conserving benchmarks.   It may  also be
interesting to extend  this approach to a wider  class of CP-violating
models  within the  general MSSM  framework.  Since  it has  many more
CP-violating parameters,  our approach  may be a  useful guide  to the
possibilities opened  up by this larger parameter  space.  Needless to
say, our approach  may also be interesting for  other scenarios for CP
violation,  beyond  the MSSM  and  indeed  supersymmetry. Finally,  we
observe  that this  geometric  approach is  not  restricted to  issues
related only to  CP violation. It could find  broader applicability to
other problems  where one wants  to maximize observables subject  to a
set of constraints.

\vspace{-0.2cm}
\subsection*{Acknowledgements}
\vspace{-0.3cm}
\noindent
We thank Maxim Pospelov, Adam Ritz and We-Fu Chang for helpful discussions.
The work of  A.P. was supported  in part by  the STFC
research grant PP/D000157/1.

\newpage

\def\theequation{\Alph{section}.\arabic{equation}}
\begin{appendix}

\setcounter{equation}{0}
\section{Two-Loop Gaugino Contributions to EDMs}

In this  Appendix we calculate  a particular set of  EDM contributions
induced by two-loop diagrams involving chargino ($\tilde\chi^\pm$) and
neutralino ($\tilde\chi^0$) quantum  effects.  The relevant tree-level
interactions are:
\begin{eqnarray}
{\cal L}_{A\bar{f}f}&=& -\,e\,  \overline{\tilde\chi^+}\gamma^\mu
\tilde\chi^+\, A_\mu\ 
=\ +\,e\,  \overline{\tilde\chi^-}\gamma^\mu \tilde\chi^-\, A_\mu\; ,
\nonumber \\[0.2cm]
{\cal L}_{H^\pm\tilde\chi_i^0\tilde\chi_j^\mp}&=& -\frac{g}{\sqrt{2}}\,
H^+\,\overline{\tilde\chi_i^0}\,\left(
g^S_{H^+\tilde\chi_i^0\tilde\chi_j^-} + i \gamma_5\,
g^P_{H^+\tilde\chi_i^0\tilde\chi_j^-}
\right)\, \tilde\chi_j^- \ + \ {\rm H.c.}\; ,
\nonumber \\[0.2cm]
{\cal L}_{W^\pm\tilde\chi_i^0\tilde\chi_j^\mp}&=& -\frac{g}{\sqrt{2}}\,
W^+_\mu\,\overline{\tilde\chi_i^0}\,\gamma^\mu\,\left(
g^L_{W^+\tilde\chi_i^0\tilde\chi_j^-} P_L +
g^R_{W^+\tilde\chi_i^0\tilde\chi_j^-} P_R
\right)\, \tilde\chi_j^- \ + \ {\rm H.c.}\; ,
\end{eqnarray}
where  the $H^\pm$-boson  couplings are  given in  the  {\tt CPsuperH}
manual~\cite{cpsuperh}, and the $W^\pm$-boson couplings are given by
\begin{eqnarray}
g^L_{W^+\tilde\chi_i^0\tilde\chi_j^-} &=&
N_{i3} (C_L)^*_{j2} +\sqrt{2} N_{i2} (C_L)^*_{j1}\; ,
\nonumber \\
g^R_{W^+\tilde\chi_i^0\tilde\chi_j^-} &=&
-N_{i4}^* (C_R)^*_{j2} +\sqrt{2} N_{i2}^* (C_R)^*_{j1}\; ,
\end{eqnarray}
in terms of chargino and neutralino mixing matrices.
The charged-Higgs couplings to the SM particles are given by
\begin{eqnarray}
{\cal L}_{H^\pm f f^\prime} = - g_{ff^\prime}
H^+\,\overline{f}\,\left(
g^S_{H^+\bar f f^\prime} + i \gamma_5\,
g^P_{H^+\bar f f^\prime}
\right)\, f^\prime \ + \ {\rm H.c.} ,
\end{eqnarray}
where
\begin{eqnarray}
g_{\nu l}=-\frac{gm_l}{\sqrt{2}M_W}\; , \qquad
g^S_{H^+\bar \nu l} =t_\beta/2\;, \qquad
g^P_{H^+\bar \nu l} =-it_\beta/2\; .
\end{eqnarray}
We include the threshold corrections due to the exchanges of gluinos:
\begin{eqnarray}
g_{ud} &=& -\frac{gm_u}{\sqrt{2}M_W}\; , \nonumber \\
g^S_{H^+\bar u d} &=& \frac{1}{2}\,\left(
\frac{1}{t_\beta} +
\frac{t_\beta}{1+\Delta_d^*\,t_\beta}\,\frac{m_d}{m_u} \right)\;  ,
\nonumber \\
g^P_{H^+\bar u d} &=& \frac{i}{2}\,\left(
\frac{1}{t_\beta} -
\frac{t_\beta}{1+\Delta_d^*\,t_\beta}\,\frac{m_d}{m_u} \right)\; , 
\end{eqnarray}
where
\begin{equation}
\Delta_d =\frac{2 \alpha_s}{3\pi} \mu^* M_3^*
I(M_{\tilde{D}}^2,M_{\tilde{Q}}^2,|M_3|^2)\; . 
\end{equation}
We note  that this  threshold correction induces  a dependence  on the
gluino  mass phase  $\Phi_3$.  At  the two-loop  level,  charginos and
neutralinos     induce     CP-violating     interactions    in     the
$\gamma$-$H^\pm$-$W^{\mp}$   and  $\gamma$-$W^\pm$-$W^\mp$  couplings,
which in turn produce non-zero electron and quark EDMs.

\subsection{EDM Induced by the {\boldmath $\gamma$-$H^\pm$-$W^{\mp}$} Coupling}

In addition to squarks~\cite{APLB}, charginos and neutralinos may also
induce a complex and CP-violating effective $\gamma$-$H^\pm$-$W^{\mp}$
coupling~\cite{Li:2008kz}.       The       CP-violating      effective
$\gamma$-$H^\pm$-$W^{\mp}$ coupling can then give rise to electron and
$d$-quark EDMs.  Specifically, defining  as $f^\prime \equiv  l,d$ all
fermions with weak isospin $T_z = -1/2$, we have
\begin{eqnarray}
\left(\frac{d^E_{f^\prime}}{e}\right)^{WH^\pm} &= &
\frac{\alpha^2}{64\pi^2 s_W^4}\,
\left(\frac{-\sqrt{2}\, g_{ff^\prime}}{g}\right)\,
\frac{1}{M_{H^\pm}^2}\,
\nonumber \\
&& \hspace{-2.0cm}
\times \sum_{i=1}^4 \sum_{j=1}^2 \Bigg\{ \int {\rm d}x\, \frac{1}{1-x}\,
J\left(r_{WH^\pm},\frac{r_{jH^\pm}}{1-x}+\frac{r_{iH^\pm}}{x} \right)
\nonumber \\
&& \Bigg[
~\imag{\left((g^S_{H^+\bar f f^\prime} + i g^P_{H^+\bar f f^\prime})\,
G^{RL}_+\right)} \,m_{\tilde\chi_j^\pm}\,x^2
\nonumber \\
&&  +
\imag{\left((g^S_{H^+\bar f f^\prime} + i g^P_{H^+\bar f f^\prime})\,
G^{LR}_+\right)} \,m_{\tilde\chi_i^0}\,(1-x)^2
\nonumber \\
&&  +
\imag{\left((g^S_{H^+\bar f f^\prime} + i g^P_{H^+\bar f f^\prime})\,
G^{RL}_-\right)} \,m_{\tilde\chi_j^\pm}\,x
\nonumber \\
&&  +
\imag{\left((g^S_{H^+\bar f f^\prime} + i g^P_{H^+\bar f f^\prime})\,
G^{LR}_-\right)} \,m_{\tilde\chi_i^0}\,(1-x) \Bigg] \Bigg\}\; ,
\end{eqnarray}
where $r_{xy} \equiv M_x^2/M_y^2$ and
\begin{eqnarray}
G^{AB}_{\pm} &\equiv & ~
\left(g^S_{H^+\tilde\chi_i^0\tilde\chi_j^-}\right)^*
\left(g^A_{W^+\tilde\chi_i^0\tilde\chi_j^-} \pm
g^B_{W^+\tilde\chi_i^0\tilde\chi_j^-} \right)
\nonumber \\
& +  &
i\,\left(g^P_{H^+\tilde\chi_i^0\tilde\chi_j^-}\right)^*
\left(g^A_{W^+\tilde\chi_i^0\tilde\chi_j^-} \mp
g^B_{W^+\tilde\chi_i^0\tilde\chi_j^-} \right)\; , 
\end{eqnarray}
with $A,B = L,R$. The loop function $J(a,b)$ is defined
in terms of the function $J(x)$,
\begin{equation}
  \label{Jab}
J(x)\  \equiv \ \frac{x\ln x}{x-1}\; ,
\end{equation}
by $J(a,b)  \equiv [J(a)-J(b)]/(a-b)$.  For $a= b$,  the loop function
$J(a,b)$ takes  on the simple form: $J(a,a)=(-\ln a + a-1)/(a-1)^2$,
with $J(1,1)=1/2$.
Observe that the  expression $g^S_{H^+\bar f f^\prime}+ig^P_{H^+\bar f
f^\prime} =  t_\beta\,(t_\beta\, m_d/m_u)$  for $f^\prime =  l~(d)$ is
real  at the  tree  level, but  becomes  in general  complex when 
gluino threshold corrections are included, as discussed above.

\subsection{EDM Induced by the {\boldmath $\gamma$-$W^\pm$-$W^\mp$} Coupling}

Quantum  loops of  charginos and  neutralinos generate  P-  and CP-odd
interactions          in          the         $\gamma$-$W^\pm$-$W^\mp$
coupling~\cite{THW,Chang:2005ac,Giudice:2005rz}, which in turn produce
non-zero fermion EDMs at the  two-loop level.  The analytic results we
use  are  based on  the  latest calculation  in~\cite{Giudice:2005rz}.
In~detail,  if we  define  $f  \equiv l,d$,  the  contribution of  the
$\gamma$-$W^\pm$-$W^\mp$ coupling to the $f$-particle EDM is given by
\begin{equation}
\left(\frac{d^E_{f}}{e}\right)^{WW} =
\frac{\alpha^2}{32\pi^2 s_W^4}\,
\imag{\left[g^L_{W^+\tilde\chi_i^0\tilde\chi_j^-}
\left(g^R_{W^+\tilde\chi_i^0\tilde\chi_j^-}\right)^*\right]}\,
\frac{m_f\, m_{\tilde\chi_i^0}\, m_{\tilde\chi_j^\pm}}{M_W^4}\,
f_{WW}(r_i,r_j)\; ,
\end{equation}
where
\begin{equation}
f_{WW}(r_i,r_j)=\int_0^1\,\frac{{\rm d}x}{1-x}\,
J\left(0\,,\frac{(1-x)r_i+x r_j}{x(1-x)}\right)\; ,
\end{equation}
with
$r_j\equiv m_{\tilde\chi_j^\pm}^2/M_W^2$ and
$r_i\equiv m_{\tilde\chi_i^0}^2/M_W^2$, and $J(a,b)$ being defined
after~(\ref{Jab}).

\subsection{ {\tt CPsuperH2.2} Interface}

The  additional   two-loop  EDMs   discussed  above  have    been
implemented in the public  version of the {\tt CPsuperH2.2} code.  The output
of the  two-loop EDM calculations and the strange-quark  chromo-EDM is contained
in  the auxiliary  array {\tt  RAUX\_H}.  The  additional two-loop
EDMs  involving  charginos and  neutralinos  have  been  added to  the
corresponding EDMs presented  in Ref.~\cite{ELPEDM} to yield the
total EDMs. Specifically, the following assignment of variables (all in
units  of  ${\rm  cm}$)  has  been  made  in  the  updated  code  {\tt
CPsuperH2.2}:
\begin{itemize}
\item Two-loop electron EDMs  induced by the $\gamma$-$H^\pm$-$W^{\mp}$
and $\gamma$-$W^\pm$-$W^\mp$ couplings:
\begin{eqnarray}
{\tt RAUX\_H(205)} &=& (d^E_e/e)^{W H^\mp}\,, \ \ \
{\tt RAUX\_H(206)}  =  (d^E_e/e)^{W W}\,.
\end{eqnarray}
\item Two-loop down-quark EDMs induced by the $\gamma$-$H^\pm$-$W^{\mp}$
and $\gamma$-$W^\pm$-$W^\mp$  couplings:
\begin{eqnarray}
{\tt RAUX\_H(225)} &=& (d^E_d/e)^{W H^\mp}\,, \ \ \
{\tt RAUX\_H(226)}  =  (d^E_d/e)^{W W}\,. 
\end{eqnarray}
\item The corresponding two-loop strange-quark EDMs:
\begin{eqnarray}
{\tt RAUX\_H(235)} &=& (d^E_s/e)^{W H^\mp}\,, \ \ \
{\tt RAUX\_H(236)}  =  (d^E_s/e)^{W W}\,.
\end{eqnarray}
\item The chromo-EDM of the $s$-quark:
\begin{equation}
{\tt RAUX\_H(400)} = d^C_s = (d^C_s)^{\tilde{\chi}^\pm}
+(d^C_s)^{\tilde{\chi}^0}+(d^C_s)^{\tilde{g}}
+(d^C_s)^{H} \;,
\end{equation}
where the individual contributions are given by
\begin{eqnarray}
{\tt RAUX\_H(401)} &=&(d^C_s)^{\tilde{\chi}^\pm}\,, \ \ \
{\tt RAUX\_H(402)}  = (d^C_s)^{\tilde{\chi}^0}\, ,\nonumber \\
{\tt RAUX\_H(403)} &=&(d^C_s)^{\tilde{g}}\,, \ \ \ \ \
{\tt RAUX\_H(404)}  = (d^C_s)^{H}\,.
\end{eqnarray}
\end{itemize}

\end{appendix}

\end{document}